\documentclass[11pt]{article}
\usepackage{jheppub} 
\usepackage[utf8]{inputenc}
\usepackage[english]{babel}
\usepackage{amsmath}
\usepackage{subfigure}
\usepackage{bigints}
\usepackage{color}
\usepackage{amsfonts}
\usepackage{amssymb}
\usepackage{graphicx}
\newcommand{\be}{\begin{equation}}
\newcommand{\ee}{\end{equation}}
\newcommand{\bea}{\begin{eqnarray}}
\newcommand{\eea}{\end{eqnarray}}

\newcommand{\zmt}{Z'}

\newcommand{\nn}{\nonumber}

\catcode`@=12

\begin{document}

\title{FIMP dark matter candidate(s) in a $B-L$ model with inverse seesaw mechanism}

\author[1,2]{Waleed Abdallah,}
\author[1,3]{Sandhya Choubey}
\author[1,4]{and Sarif Khan}
\affiliation[1]{Harish-Chandra Research Institute, HBNI, Chhatnag Road,
Jhunsi, Allahabad 211019, India}
\affiliation[2]{Department of Mathematics, Faculty of Science, Cairo University, Giza 12613, Egypt}
\affiliation[3]{Department of Physics, School of
Engineering Sciences, KTH Royal Institute of Technology, AlbaNova
University Center, 106 91 Stockholm, Sweden}
\affiliation[4]{Institut f\"{u}r Theoretische Physik,
Georg-August-Universit\"{a}t G\"{o}ttingen, Friedrich-Hund-Platz 1,
G\"{o}ttingen, D-37077 Germany}
\emailAdd{waleedabdallah@hri.res.in,
          sandhya@hri.res.in, sarifkhan@hri.res.in}
\abstract{
The non-thermal dark matter (DM) production via the so-called freeze-in mechanism provides a simple alternative to the standard thermal WIMP scenario. In this work, we consider a popular $U(1)_{B-L}$ extension of
the standard model (SM) in the context of inverse seesaw mechanism which has {\it at least} one (fermionic) FIMP DM candidate. Due to the added $\mathbb{Z}_{2}$ symmetry, a SM gauge singlet fermion, with mass of order keV, is stable and can be a warm DM candidate. Also, the same $\mathbb{Z}_{2}$ symmetry helps the lightest right-handed neutrino, with mass of order GeV, to be a stable or long-lived particle by making a corresponding Yukawa coupling very small. This provides a possibility of a two component DM scenario as well. Firstly, in the  absence of a GeV DM component ({\it i.e.}, without tuning its corresponding Yukawa coupling to be very small), we consider only a keV DM as a single component DM, which is produced by the freeze-in mechanism via the decay of the extra $Z'$ gauge boson associated to $U(1)_{B-L}$ and can consistently explain the DM relic density measurements. In contrast with most of the existing literature, we have found a reasonable DM production from the annihilation processes. 
After numerically studying the DM production, we show the dependence of the DM
relic density as a function of  its relevant free parameters. We use these results to obtain the parameter space regions that are compatible with the DM relic density bound. Secondly, we study
a two component DM scenario and emphasize that the current DM relic density bound can be satisfied for a wide range of parameter space.
}
\maketitle
\section{Introduction}\label{Intro}
The standard model (SM) is a very successful theory in describing nature.
The discovery of the last missing piece of the SM, {\it viz.}, the Higgs boson,
further increases its concreteness. In spite of its tremendous success, the SM can not
explain a number of phenomena - two of the most important ones being the presence
of dark matter (DM) and non-zero neutrino mass. Presence of DM in the universe
is a very well established fact. The first indication of DM came from the observation of
Galactic velocities within the Coma cluster by Fritz Zwicky 
in 1933~\cite{Zwicky:1933gu}, followed by the observation
of galaxy rotation curves by Vera Rubin in 1970~\cite{Rubin:1970zza}.
Subsequently, the observation of bullet
cluster~\cite{Clowe:2006eq} firmly confirmed the presence of DM. Currently the 
best measurement of the amount of DM present in the universe comes from the 
Planck data~\cite{Ade:2015xua},
\begin{eqnarray}
\Omega\, h^2 = 0.1199 \pm 0.0027\,\, {\rm at\,\,68\%\,\,CL}\,,
\end{eqnarray}  
where $h$ is the reduced Hubble parameter and of order unity.
Unfortunately, the SM does not have any fundamental particle  which can be a viable DM candidate. Therefore, to address
the issue of DM from particle physics point of view, we need to extend the
SM particle content and/or its gauge group. One of the most promising scenarios is to consider the DM candidate as a Weakly Interacting Massive Particle (WIMP) \cite{Gondolo:1990dk,Srednicki:1988ce}, which is produced in the early universe through the thermal freeze-out mechanism~\cite{Gondolo:1990dk,Srednicki:1988ce}.  
However, WIMP type DM attracts stringent bounds from direct and indirect detection experiments~\cite{Aprile:2018dbl,Tan:2016zwf,Akerib:2016vxi,Amole:2017dex,Aprile:2019dbj,Ahnen:2016qkx,Abdallah:2016ygi,Giesen:2015ufa}. 
In particular, a large portion of the parameter space in the
spin independent/dependent WIMP-nucleon cross section and DM mass plane
is ruled out by the direct detection (DD) bounds. Moreover, 
in near future with increasing sensitivity of the DD experiments \cite{Aprile:2018dbl,Tan:2016zwf,Akerib:2016vxi,Amole:2017dex,Aprile:2019dbj}, these bounds might touch the so-called neutrino floor \cite{Drukier:1983gj,Cushman:2013zza}. In this work, we follow a non-thermal way of DM
production, {\it viz.}, via the freeze-in mechanism \cite{Hall:2009bx}. In this scenario, the 
DM is very feebly interacting with the other particles, and as a result 
never achieves thermal equilibrium in the early universe with the cosmic soup. 
Hence it is named Feebly Interacting Massive Particles (FIMPs). Due to their very feeble interactions, FIMPs easily escape the above mentioned DD  bounds while satisfying the measured value for the DM relic density~\cite{Hall:2009bx, Konig:2016dzg, Biswas:2016bfo, Biswas:2016iyh, Biswas:2016yjr, Biswas:2017tce, Biswas:2017ait,  Kaneta:2016vkq,Caputo:2018zky}. 

On the other hand,  results of the neutrino oscillation experiments
\cite{Cowan:1992xc, Fukuda:1998mi, Ahmad:2002jz, Eguchi:2002dm, An:2015nua, RENO:2015ksa, Abe:2014bwa, Abe:2015awa, Salzgeber:2015gua, Adamson:2016tbq, Adamson:2016xxw} have confirmed oscillations between neutrino flavours. Since neutrino flavour oscillations are a clear proof of the neutrinos being massive and mixed, the neutrino oscillation experiments contradict the SM which postulates that the neutrinos are massless. Consequently, in order to explain tiny neutrino masses, one has to extend the SM
by adding new particles and/or additional gauge groups.

In the present work we explain the above two 
puzzles by extending the SM gauge group by a $U(1)_{B-L}$ gauge symmetry as a simple (minimal) and well motivated extension of the SM, where $B$ is the baryon number and $L$ is the lepton number.
In addition to the extra neutral gauge boson $Z'$ associated with the $U(1)_{B-L}$, an extra SM singlet scalar $\phi_H$ (charged under $U(1)_{B-L}$ to break $B-L$ gauge symmetry spontaneously) is added in this simple extension, which leads to interesting signatures at the LHC \cite{Khalil:2006yi,Emam:2007dy,Huitu:2008gf,Blanchet:2009bu,Basso:2008iv}. Moreover, nine additional SM singlet fermions ($N_R^i$ and $ S^i_{1,2},~i=1,2,3$) are needed to explain the naturally\footnote{Here, ``naturally" means the Dirac neutrino masses, $M_D$, have the same size as the Dirac masses of the SM fermions and, in contrary to the usual type-I seesaw mechanism \cite{Mohapatra:1980qe,Marshak:1979fm,Wetterich:1981bx,Masiero:1982fi,Mohapatra:1982xz,Buchmuller:1991ce,Abbas:2007ag}, large Dirac neutrino Yukawa couplings, $\lambda_d \sim {\cal O}(0.1)$,  with right-handed neutrino ($N_R^i$) masses are of order TeV. } small neutrino masses through the inverse seesaw mechanism \cite{MohapatraIS1,Mohapatra:1986bd,Khalil:2010iu,GonzalezGarcia:1988rw}. These additional fermions are not only required to
generate the tiny neutrino masses via the inverse seesaw mechanism but are also
needed for the gauge anomaly cancellation. In such a framework, three of these SM singlet fermions, $S_1^i$, are completely decoupled due to the introduction of $\mathbb{Z}_2$ symmetry and have naturally small mass (of order keV) according to 't Hooft's naturalness criterion~\cite{tHooft:1979rat}. Therefore, the lightest one, $S^1_1$, will be a stable particle and hence a warm DM (WDM) candidate~\cite{Pagels:1981ke, Peebles:1982ib, Bond:1982uy, Olive:1981ak}, as discussed in Ref.~\cite{El-Zant:2013nta}.
Moreover, since these keV mass singlet fermions are odd under the $\mathbb{Z}_2$ symmetry, they have no mixing with the active
neutrinos and consequently are safe from the bound imposed by the x-ray observations
\cite{Boyarsky:2009ix}. In Ref.~\cite{El-Zant:2013nta}, an extra moduli field was introduced to produce 
this keV WDM non-thermally to achieve the
correct ballpark value of relic density consistent with the WMAP and Planck observations. In the current work, {\it without} introducing any extra field contrary to Ref.~\cite{El-Zant:2013nta}, we successfully produce the keV WDM by the freeze-in
mechanism through the decay and annihilation channels of $Z'$. After explaining the
keV FIMP WDM as a successful single component FIMP DM scenario to satisfy the correct value of the DM relic density, we study a two component FIMP DM as another possible scenario in the present model, where in addition to the FIMP WDM $S_1^1$, the lightest heavy right-handed neutrino $\nu_H^1$ can be a  FIMP DM (with mass of order GeV) by tuning its corresponding Yukawa coupling to be very small \cite{Fiorentin:2016avj,DiBari:2016guw}.
The GeV scale FIMP DM can be produced through the decay and annihilation processes of both the extra neutral gauge boson $Z'$
as well as the extra $B-L$ Higgs $h'$, while the keV FIMP WDM is produced only through the decay and annihilation processes of $Z'$.

The rest of the paper is organized as follows. In section~\ref{BLSMIS} we discuss the $B-L$ model with inverse seesaw mechanism and how the light neutrinos acquire their tiny masses. In section~\ref{WDMFIMP} we show that a keV sterile neutrino can be a WDM and produce the observed DM relic density as a single component FIMP DM. Section~\ref{2DM} is dedicated for
studying two component FIMP type DM. Finally, our conclusions are given in section~\ref{Con}.

\section{$B-L$ model with inverse seesaw scenario and neutrino masses}\label{BLSMIS}
The gauged $B-L$ extension of the SM (BLSM) is based on the gauge group  $SU(3)_C\times
SU(2)_L\times U(1)_Y\times U(1)_{B-L}$. By imposing $U(1)_{B-L}$, the gauge sector of the SM is extended to include a new neutral gauge boson $Z'$ associated with the $B-L$ gauge symmetry. In addition, it has three SM singlet fermions $N_R^i$ (three right-handed neutrinos) with $B-L$ charge $=-1$ that arise as a result of anomaly cancellation conditions. Included also is an extra SM singlet scalar $\phi_H$ with $B-L$ charge $=-1$, while $\phi_h$ is the usual electroweak (EW) Higgs doublet. In order to satisfy the experimental measurements for the non-vanishing light neutrino masses
with TeV scale right-handed (RH) neutrino using type-I seesaw mechanism, a very small Dirac neutrino Yukawa couplings, $\lambda_d \lesssim {\cal O}(10^{-6})$ must be assumed \cite{Mohapatra:1980qe,Marshak:1979fm,Wetterich:1981bx,Masiero:1982fi,Mohapatra:1982xz,Buchmuller:1991ce,Abbas:2007ag}. Therefore, the mixing angle between the left- and right-handed neutrinos is quite suppressed, as it is proportional to $\lambda_d \lesssim {\cal O}(10^{-6})$. As a consequence of such small mixing angle, the interactions between the RH neutrinos and the SM particles are very suppressed, making it difficult to
observe them at the LHC \cite{Khalil:2006yi,Emam:2007dy,Huitu:2008gf,Blanchet:2009bu,Basso:2008iv}. Thus, we generate neutrino masses using 
the so-called inverse seesaw mechanism \cite{MohapatraIS1,Mohapatra:1986bd,Khalil:2010iu,GonzalezGarcia:1988rw}
that can naturally accommodate light neutrino masses with TeV scale RH neutrinos and large Yukawa couplings. In addition to  the particle content as mentioned above, the BLSM with Inverse Seesaw (BLSMIS) has three extra pairs of SM singlet fermions $(S_{1,2}^i,~i =1,2,3)$ with $B-L$ charge $=\mp 2$, respectively.
In Table~\ref{tab1}, we show the complete particle spectrum for the BLSMIS model 
with their associated charges for different gauge groups. An additional discrete symmetry has been introduced, {\it viz.}, $\mathbb{Z}_{2}$. All BLSMIS particles are even under this symmetry except $S_1$ which is odd. Due to this symmetry, terms like $N^c_{R} \phi^\dagger_{H} S_{1}$ and $S_1 S_2$, that could spoil the usual inverse seesaw mechanism, are forbidden \cite{MohapatraIS1,Mohapatra:1986bd,Khalil:2010iu,GonzalezGarcia:1988rw}. The complete Lagrangian for
this model is given~by
\def\I{i}
\begin{table}[t!]
\begin{center}
\small
\begin{tabular}{||@{\hspace{0cm}}c@{\hspace{0cm}}|@{\hspace{0cm}}c@{\hspace{0cm}}|@{\hspace{0cm}}c@{\hspace{0cm}}|@{\hspace{0cm}}c@{\hspace{0cm}}||}
\hline
\hline
\begin{tabular}{c}
    Gauge\\
    Group\\ 
    \hline
    
    $SU(2)_{L}$\\ 
    \hline
    $U(1)_{Y}$\\ 
    \hline
    $U(1)_{B-L}$\\ 
\end{tabular}
&
\begin{tabular}{c|c|c}
    \multicolumn{3}{c}{Baryon Fields}\\ 
    \hline
    $Q_{L}^{i}=(u_{L}^{i},d_{L}^{i})^{T}$&$u_{R}^{i}$&$d_{R}^{i}$\\ 
    \hline
    $2$&$1$&$1$\\ 
    \hline
    $1/6$&$2/3$&$-1/3$\\ 
    \hline
    $1/3$&$1/3$&$1/3$\\ 
\end{tabular}
&
\begin{tabular}{c|c|c|c|c}
    \multicolumn{5}{c}{Lepton Fields}\\
    \hline
    $L_{L}^{i}=(\nu_{L}^{i},e_{L}^{i})^{T}$ & $e_{R}^{i}$ & $N_{R}^{i}$ & $S_{1}^{i}$ & $S_{2}^{i}$\\
    \hline
    $2$&$1$&$1$&$1$&$1$\\
    \hline
    $-1/2$&$-1$&$0$&$0$&$0$\\
    \hline
    $-1$&$-1$&$-1$&$-2$&$2$\\
\end{tabular}
&
\begin{tabular}{c|c}
    \multicolumn{2}{c}{Scalar Fields}\\
    \hline
    ~~~$\phi_{h}$~~~&$\phi_{H}$\\
    \hline
    $2$&$1$\\
    \hline
    $1/2$&$0$\\
    \hline
    $0$&$-1$\\
\end{tabular}\\
\hline
\hline
\end{tabular}
\caption{Complete particle spectrum and their corresponding
charges under various gauge groups.}
\label{tab1}
\end{center}    
\end{table}
\begin{eqnarray}\label{Lag}
\mathcal{L} &=& \mathcal{L}_{\rm SM} 
- \frac{1}{4} F^\prime_{\mu\nu} F^{\prime\,\mu\nu}
+ (D_{\mu} \phi_{H})^{\dagger} D_{\mu} \phi_{H}
+ \frac{i}{2} \bar{N_{R}} \gamma^{\mu} D_{\mu} N_{R}
+ \frac{i}{2} \bar{S_{1}} \gamma^{\mu} D_{\mu} S_{1} 
+ \frac{i}{2} \bar{S_{2}} \gamma^{\mu} D_{\mu} S_{2} \nn \\
&-& \mathcal{V}(\phi_h, \phi_{H})- (\lambda_{d} \bar{L} \tilde{\phi_h} N_R 
+ \lambda_s \bar{N}^c_{R} \phi_{H} S_{2} + {\it h.c.}),
\end{eqnarray}  
where $F^\prime_{\mu\nu}=\partial_\mu B'_{\nu}-\partial_\nu B'_{\mu}$ is the $U(1)_{B-L}$ field strength, $D_{\mu}$ is the covariant derivative, $\tilde{\phi_{h}} = i \sigma_2 \phi_h$ and the flavor indices are omitted for simplicity. The general structure
of the covariant derivative $D_{\mu}$ in the present model takes the following
form
\begin{eqnarray}
D_{\mu} = \partial_{\mu} - i g_c T^{\alpha} G^{\alpha}_{\mu} - i g \tau^a W^a_{\mu}
-i g_Y Y B_{\mu} - i g' Y_{BL} B'_{\mu}\,,
\end{eqnarray}
where $(g_c, T^{\alpha}, G^{\alpha}_{\mu})$ are the $SU(3)_C$ gauge coupling,
generator and the gauge field, respectively. Similarly, $(g, \tau^{a}, W^a_{\mu})$,
$(g_{Y}, Y, B_{\mu})$ and $(g', Y_{BL}, B'_{\mu})$ are the corresponding
quantities for $SU(2)_L$, $U(1)_Y$ and $U(1)_{B-L}$, respectively. It is worth mentioning that a kinetic mixing term $F^\prime_{\mu\nu} F^{\mu\nu}$ is allowed  and it leads to a non-vanishing $Z$-$Z'$ mixing angle, $\theta'$ \cite{Holdom:1985ag,Chankowski:2006jk,Krauss:2012ku}. However, due to the stringent constraint from LEP experiments on the $Z$-$Z'$ mixing angle ($|\theta'|\lesssim 10^{-3}$) \cite{Abreu:1994ria,Alcaraz:2006mx,Erler:2009jh}, one may neglect this term. Finally, the potential $\mathcal{V}(\phi_h,\phi_H)$
is given by~\cite{Emam:2007dy,Khalil:2006yi}  
\begin{eqnarray}
\mathcal{V}(\phi_h,\phi_{H}) = \mu_{h}^2~\phi_h^\dagger \phi_h + \mu_{H}^2~\phi_H^\dagger \phi_H + \lambda_h (\phi_h^\dagger \phi_h)^2  + \lambda_H (\phi_H^\dagger \phi_H)^2
+ \lambda_{hH} (\phi_h^\dagger \phi_h) (\phi_H^\dagger \phi_H)\,,
\end{eqnarray}
where the potential $\mathcal{V}(\phi_h,\phi_{H})$ will be bounded from below when the following inequalities are satisfied simultaneously 
\begin{eqnarray}
\mu^2_{h} < 0,\,\,\,\mu^2_{H} < 0,\,\,\,\lambda_{h} \geq 0,\,\,\,\lambda_H \geq 0\,\,{\rm and}\,\,\, \lambda_{hH} \geq -2 \sqrt{\lambda_h\lambda_H}\,.
\end{eqnarray}
Here, both the scalars $\phi_H$ and $\phi_h$ acquire their non-zero vacuum expectation values (VEVs), therefore, the $B-L$ and the EW symmetries are broken spontaneously
and the SM Higgs doublet $\phi_h$ and the $B-L$ singlet $\phi_H$ take the following form:
\begin{eqnarray}
\phi_{h}=
\begin{pmatrix}
0 \\
\dfrac{v+h}{\sqrt{2}}
\end{pmatrix},
\,\,\,\,\,\,\,\,\,
\phi_{H}=
\dfrac{v'+H}{\sqrt{2}}\, ,
\label{phih}
\end{eqnarray} 
where $v \simeq$ 246 GeV is the EW symmetry breaking scale and $v'$ is
the scale of $B-L$ symmetry breaking which is, in general, unknown and ranging from TeV to much higher scales. After breaking the $B-L$ and the EW symmetries spontaneously, 
the extra neutral gauge boson $Z'$ acquires its mass $(M_{Z'}=g' v')$~\cite{Emam:2007dy,Khalil:2006yi}\footnote{The experimental search for $Z'$, by LEP~II \cite{Cacciapaglia:2006pk,Carena:2004xs}, leads to another constraint: $M_{Z'}/g'\gtrsim 7$~TeV. This constraint will easily be satisfied due to a smallness of $g'$ which is required by the freeze-in scenario \cite{Hall:2009bx}.}, and the neutrino Yukawa interaction terms in Eq.~(\ref{Lag}) and in addition a very small Majorana mass $\mu_S$ for $S_{1,2}$ lead to the following neutrino mass terms\footnote{$\mu_S$ is naturally small due to 't Hooft's naturalness criterion \cite{tHooft:1979rat}, for simplicity we assume $S_1$ and $S_2$ have the same small Majorana mass ($\mu_S$), and the generation of such small
$\mu_S$ from non-renormalizable terms has been discussed in \cite{Abdallah:2011ew} and radiatively in \cite{Ma:2009gu}.}
\begin{eqnarray}
\mathcal{L}^{\nu}_{m} = \mu_{S}(\bar{S}^c_1 S_1+\bar{S}^c_2 S_2)+(M_{D} \bar{\nu}_{L} N_{R} + M_{N} \bar{N}^c_R S_2 + {\it h.c.}),
\end{eqnarray}
where $M_{D} = \lambda_{d} v/\sqrt{2}$ and $M_{N} = \lambda_{s} v'/\sqrt{2}$.
Therefore, the neutrino mass matrix in the basis $(\nu^c_{L}, N_{R}, S_2, S_1)$ can be written as 
\begin{eqnarray}
\mathcal{M}_{\nu} = \left(\begin{array}{cccc}
0 ~~&~~ M_{D} ~~&~~ 0 ~~&~~ 0\\

M_{D}^{T} ~~&~~ 0 ~~&~~ M_{N} ~~&~~ 0\\

0 ~~&~~ M^T_{N} ~~&~~ \mu_{S} ~~&~~ 0\\

0 ~~&~~ 0 ~~&~~ 0 ~~&~~ \mu_{S}
\end{array}\right).
\label{neutrino-mass}
\end{eqnarray}
It is clearly seen that $S_1$ is completely decoupled and has no mixing with active neutrinos. It only interacts with the neutral gauge boson $Z'$ with a coupling $g_{Z'S_1S_1}=g'$. Therefore, $S_1$ is free from cosmological and astrophysical constraints coming from active-sterile mixing  \cite{Boyarsky:2009ix}. Thus its mass is given as,
\begin{equation}
M_{S_1} = \mu_{S}\,.
\end{equation}
After diagonalising the upper left $3\times 3$ submatrix of the neutrino mass matrix $\mathcal{M}_{\nu}$, the light and heavy neutrino masses, respectively, are given by  
\begin{eqnarray}
M_{\nu_l} & \simeq & M_{D} M_{N}^{-1} \mu_{S} (M_{N}^T)^{-1} M_{D}^T \,, \\
M^2_{\nu_{H,H'}} &\simeq& M^2_{N} + M^2_{D} \mp \frac{1}{2} \frac{M_N^2 \mu_{S}}{M_N^2+M_D^2}\,,\end{eqnarray} 
where $\mu_{S}\ll M_D,M_N$ is assumed. One can naturally obtain eV scale light neutrino masses
with $\mu_{S}$ of order keV and $M_N$ of order TeV, keeping Yukawa coupling $\lambda_d$ of order one. Such large couplings between the heavy RH neutrinos and the SM particles leads to interesting implications and enhances the accessibility of TeV scale $B-L$ at the LHC \cite{Abdelalim:2014cxa,Bandyopadhyay:2012px,Abdallah:2012nm}.  

Recall that due to the added $\mathbb{Z}_{2}$ symmetry, $S_1$ is completely decoupled. Hence the lightest fermionic singlet, $S_1^1$, is a stable particle and hence a DM candidate. Since 
 its mass ($=\mu_{S}$) is of order keV, hence $S_1^1$ is a warm DM (WDM) candidate \cite{El-Zant:2013nta}\footnote{The contribution of the new light degrees of freedom ($S_1^i$) to the number of effective neutrino species, $N_{\rm eff}$, has been checked using Eq.~(5) in Ref.~\cite{Hooper:2011aj} to calculate extra effective neutrino species, $\Delta N_{\rm eff}$, and found it to be negligible.}. Moreover, one can easily make the lightest heavy RH neutrino, $\nu_H^1$, stable or long-lived by taking the corresponding Yukawa coupling to be very small $\lesssim 3\times 10^{-26} ({\rm GeV}/M_N)^{1/2}$~\cite{Fiorentin:2016avj,DiBari:2016guw}. Thus, from here onwards we focus on the
two component DM scenario, where, one of them is GeV scale DM, $\nu_H^1$, and the other is 
 keV scale WDM, $S_1^1$.   

It is important to note that due to the mixing term in the potential $\mathcal{V}(\phi_h,\phi_{H})$, the squared mass matrix of the neutral Higgs bosons in the basis $(h,H)$ is non-diagonal and takes the following form:
\begin{eqnarray}
\mathcal{M}^2_{\rm scalar} = \left(\begin{array}{cc}
2\lambda_h v^2 ~&~ \lambda_{hH}\,v'\,v \\
\lambda_{hH}\,v'\,v ~&~ 2 \lambda_H v'^2
\end{array}\right).
\label{mass-matrix}
\end{eqnarray}
Rotating this matrix into the basis  $(h_1, h_2)$ which is defined as follows
\begin{eqnarray}
h_{1}&=& h \cos \alpha + H \sin \alpha \,, \nn \\
h_{2}&=& - h \sin \alpha + H \cos \alpha\,,
\end{eqnarray} 
where the mixing angle $\alpha$ takes the following form:
\begin{eqnarray}
\tan 2\alpha = \dfrac{\lambda_{hH}\,v'\,v}
{\lambda_H v'^2 - \lambda_h v^2}\,.
\end{eqnarray}
Therefore, the masses of these two physical Higgs scalars $(h_1, h_2)$ are given by\footnote{Hereafter, the physical state $h_1$ refers to the SM-like Higgs boson and its mass $M_{h_1}$ is fixed at $125.5$~GeV to agree with the LHC measurements \cite{Aad:2012tfa,Chatrchyan:2012xdj}. Also, according to the measured values of Higgs boson signal strengths for its various decay modes, the mixing angle $\alpha$ should be very small, thus we have fixed it at $0.01$~rad.}
\begin{eqnarray}
M^2_{h_{1,2}} &=& \lambda_h v^2 + \lambda_H v'^2 \mp 
\sqrt{(\lambda_H v'^2 - \lambda_h v^2)^2 + (\lambda_{hH}\,v\,v')^2}\ .
\end{eqnarray}
The quartic couplings $\lambda$'s can be written in terms of the physical masses $M_{h_{1,2}}$ as follows \cite{Basso:2010jm}
 \begin{eqnarray}
\lambda_{h}&=& \dfrac{M_{h_{1}}^{2} + M_{h_{2}}^{2} -
(M_{h_{2}}^{2} - M_{h_{1}}^{2})\cos 2 \alpha}{4\,v^{2}}\,,\nn\\
\lambda_{H} &=& \dfrac{M_{h_{1}}^{2} + M_{h_{2}}^{2} +
(M_{h_{2}}^{2} - M_{h_{1}}^{2})\cos 2 \alpha}{4 v'^2}\,,\nn \\
\lambda_{hH} &=& \dfrac{(M_{h_{2}}^{2}-M_{h_{1}}^{2})
\sin 2 \alpha}{2 v\,v'}\,.
\end{eqnarray}
We have used SARAH \cite{Staub:2013tta,Staub:2009bi,Staub:2008uz} to implement the BLSMIS and the relevant masses, couplings and decay widths have been calculated using SPheno \cite{Porod:2011nf}. 
\section{Warm DM as FIMP}\label{WDMFIMP}
As mentioned earlier, $S^1_1$ is a WDM candidate with
mass in the few keV range \cite{Adhikari:2016bei,Bernal:2017kxu,Abada:2014zra}.
We next study in detail the production of this keV DM via the freeze-in mechanism.
Here $S_1^1$ is produced solely from its coupling with the extra $U(1)_{B-L}$
gauge boson $Z'$, as mentioned in the previous section. The corresponding gauge coupling $g'$ is taken to be extremely feeble
$\sim\mathcal{O}(10^{-10})$ with the result that $S_1^1$ is never in thermal equilibrium with the cosmic soup. Due to this small $B-L$ gauge coupling, the corresponding gauge boson
$Z'$ also interacts very feebly with the cosmic soup and never attains thermal
equilibrium \cite{Arcadi:2013aba}, 
\begin{equation}
\frac{\Gamma_{Z'}}{H(T=M_{Z'})}<1,
\end{equation}
where $\Gamma_{Z'}$ is the total decay width of $Z'$ and $H$ is the Hubble parameter. Therefore, we first determine the distribution
function for $Z'$ \footnote{As $Z'$ is
not in thermal equilibrium (due to very small value of $g'$), one can not assume a Maxwell-Boltzmann distribution function for $Z'$. Therefore, the $Z'$ distribution can be found by solving
Eq.~(\ref{BME}).}.  The general formalism to determine the distribution function of any particle
(say $f$) is to solve the following Boltzmann equation:
\begin{eqnarray}
\hat{L} [f] = \mathcal{C} [f]\,,
\label{BME} 
\end{eqnarray}   
where $\hat{L}$ is the Lioville's operator and $\mathcal{C} [f]$
is known as the collision term of $f$. If we consider an isotropic and
homogeneous universe, then, using the Friedman-Robertson-Walker metric, the Lioville's operator
takes the following form:
\begin{eqnarray}\label{LE}
\hat{L} = \frac{\partial}{\partial t} - H p\, \frac{\partial}{\partial p}\,,
\end{eqnarray}
where $p$ is the absolute value of the particle's three momentum, $|\vec{p}|$. Following \cite{Konig:2016dzg}, we perform a transformation of variables, $(p,\,t)\to(\xi_p,\,z)$,
in the following way:
\begin{eqnarray}
z = \frac{M_{\rm sc}}{T}\,,\,\,\,\,\,\xi_{p} = 
\left[ \frac{g_{s}(T_0)}{g_{s}(T)}\right]^{1/3}\,\frac{p}{T}\,,
\end{eqnarray}
 where $g_{s}(T)$ is the effective entropy degrees of freedom (d.o.f)  at temperature $T$, $M_{\rm sc}$ is an arbitrary mass scale and hereafter we take it equal to the SM-like Higgs mass ($M_{\rm sc}=M_{h_1}=125.5$~GeV) and $T_0$ is the initial temperature at which the DM relic density is taken to be zero. Therefore, using the following time-temperature relation,
 \begin{equation}
\frac{d T}{d t} = -H\,T\left(1 + \frac{T g'_{s}(T)}{3 g_{s}(T)}
 \right)^{-1}\,,
 \end{equation}
 the Lioville's operator defined in Eq.~(\ref{LE}) can be simply written as
\begin{equation}
\hat{L} = z H \left(1 + \frac{T g'_{s}}{3 g_{s}}
 \right)^{-1} \frac{\partial}{\partial z}\, ,
\end{equation}
where $g'_{s}(T)$ is the derivative of $g_{s}(T)$  with respect to the temperature $T$. 

Taking only the decay term for the $Z'$
production\footnote{In principle, the collision term for
annihilation diagrams should also be considered but in this class of models those annihilation diagrams have subleading contribution \cite{Biswas:2016bfo},
hence we have not taken into account those effects and for simplicity we consider only the decay of $h_{1,2}$ as the $Z'$
production mechanism. Moreover, $h_{1,2}$ are in thermal equilibrium,
and consequently the usual equilibrium Boltzmann distribution function has been assumed
for them \cite{Konig:2016dzg}.}, the Boltzmann equation of the distribution function of $Z'$ is given by
\begin{eqnarray}
\hat{L} f_{Z'} = \sum_{i = 1,\,2} 
\mathcal{C}^{h_i \rightarrow Z' Z'} + 
\mathcal{C}^{Z' \rightarrow\, {\rm all}},
\label{Z-dis-collision}
\end{eqnarray} 
where $f_{Z'}$ is the distribution function of $Z'$, $\mathcal{C}^{h_i \rightarrow Z' Z'}$ is the collision term of $Z'$ production from the decays of scalars $h_{1,2}$
and $\mathcal{C}^{Z' \rightarrow\, {\rm all}}$ is $Z'$ decay collision term due to all its possible decay channels. The expression of these collision terms are
given in the Appendix~\ref{App:AppendixA}. Once we get the distribution function of $Z'$ by solving Eq.~(\ref{Z-dis-collision}), we then can determine
its co-moving number density by using the following relation:
\begin{eqnarray}
Y_{Z'} = \frac{n_{Z'}(z)}{s}= \frac{45 \ g }{4 \pi^4 g_{s}(M_{\rm sc}/z_0)} \int_{0}^{\infty}
d\xi_p\, \xi_p^2\, f_{Z'} (\xi_p,z)\,,
\label{number-density}
\end{eqnarray}
where $n_{Z'}$ is the $Z'$ number density, $g$ is the internal d.o.f of $Z'$ and the universe entropy density $s$ is given by $s =~(2 \pi^2/45)T^3 g_{s}(T)$ \cite{Kolb:1990vq}.

\begin{figure}[t!]
\centering
\includegraphics[angle=0,height=6.5cm,width=8.5cm]{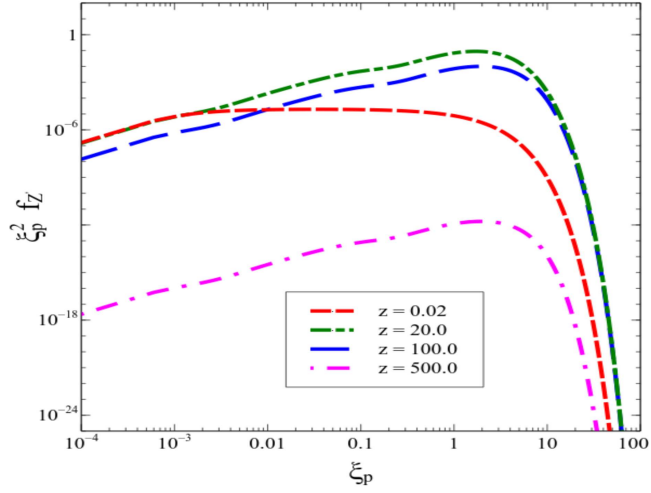}
\caption{$Z'$ distribution function versus the dimensionless
parameter $\xi_p$. Here, the relevant BLSMIS parameters are fixed as follows: $g' = 12.5 \times 10^{-10}$, $M_{Z'} = 10$~GeV,
$M_{h_2} = 1$~TeV, $M_{S_1^1} = 10$~keV, and $\alpha = 0.01$~rad.}
\label{a}
\end{figure}

From Eq.~(\ref{number-density}), one can note that the
co-moving number density of $Z'$ is directly proportional to the integrated $\xi_p^2 f_{Z'}$, {\it i.e.}, larger the area under a $\xi_p^2 f_{Z'}$ curve, larger is the $Z'$ abundance. In Fig.~\ref{a}, we show the variation of $\xi_p^2 f_{Z'}$ with respect to the dimensionless parameter $\xi_p$ for different values of $z$ ($=M_{\rm sc}/T$). As shown in the figure, areas
under the curves corresponding to $z = 0.02$ and $20.0$ are different because for higher $z = 20.0$ ({\it i.e.}, lower temperature $T$ of the universe), $Z'$ gets more time to be produced and it then subsequently decays into WDM and the SM fermions. But as $z$ is increased further (presently $z=100.0$), $Z'$ starts decaying significantly and its abundance gets depleted and the area under the curve for $z=100$ is smaller than for $z=20.0$, as seen in Fig.~\ref{a}. For still higher values of $z$ ($z = 500.0$), $Z'$ abundance decreases further due to decay. Thus, as
$z \rightarrow \infty$, $Z'$ will gradually decay to DM and its abundance eventually goes to zero. 

Once the distribution function of $Z'$ is computed, we can describe the production of
the keV DM $S^1_1$. In the present scenario, the keV DM $S_1^1$ can be produced from the decay of $Z'$, $Z' \to S_{1}^1 S_1^1$ (decay contribution), and from the annihilation processes, $f\bar{f}\to S_{1}^1 S_1^1$ mediated by $Z'$, where $f=l,q,\nu_l$. The annihilation contribution has been calculated by using  micrOMEGAs \cite{Belanger:2018mqt}. To determine the co-moving number density of the WDM $S_1^1$, we solve the
following Boltzmann equation,
\begin{eqnarray}
\frac{d Y_{S_1^1}}{dz} &=& \frac{2\, M_{\rm Pl}\, z\, \sqrt{g_{\star}}}{1.66\, M^2_{\rm sc}\, g_s}\,
\langle \Gamma_{Z' \rightarrow S_{1}^1 S_1^1} \rangle_{\rm NTH} 
\left(Y_{Z'} -  Y_{S_{1}^1} \right)\nn\\
&+&\frac{4 \pi^2}{45} \frac{M_{\rm Pl}\, M_{\rm sc} \sqrt{g_{\star}}}{1.66\, z^2}
\sum_{f}\langle {\sigma v}_{f\bar{f}\rightarrow S_{1}^1 S_1^1} \rangle
\left[\left({Y_{f}^{\rm eq}}\right)^2 -  Y^2_{S_{1}^1} \right],
\label{be-s1}
\end{eqnarray} 
where $M_{\rm Pl}=1.22\times 10^{19}$~GeV is the Planck mass, $\sqrt{g_{\star}}=\frac{g_s(z)}{\sqrt{g_{\rho}(z)}}\left(1-\frac{1}{3}\frac{{\rm d} \ln g_s(z)}{{\rm d} \ln z}\right)$, where $g_\rho$ is the effective energy degrees of freedom \cite{Gondolo:1990dk,Edsjo:1997bg}, and the non-thermal average of $Z'$ decay width is defined
by
\begin{eqnarray}
\langle \Gamma_{Z' \rightarrow S_{1}^1 S_1^1} \rangle_{\rm NTH}
= M_{Z'} \Gamma_{Z' \rightarrow S_1^1 S_1^1}\,
\frac{\bigints_{0}^{\infty} d\xi_p \ \xi_p f_{Z'}(\xi_p,z)\left(\left[\mathcal{B}(z)\frac{M_{\rm sc}}{z}\right]^2 + \left[\frac{M_{Z'}}{\xi_p}\right]^2\right)^{-1/2}}{\bigints_{0}^{\infty} d\xi_p \  \xi_p^2 f_{Z'}(\xi_p,z)}\,,
\label{non-thermal-decay}
\end{eqnarray}
where
\begin{equation}
\mathcal{B}(z) = \left[ \frac{g_{s}(M_{\rm sc}/z)}{g_{s}(M_{\rm sc}/z_0)}\right]^{1/3}.
\end{equation}
The expressions of a thermal average annihilation cross section $\langle {\sigma v}_{f\bar{f}\rightarrow S_{1}^1 S_1^1} \rangle$ and an equilibrium co-moving number density of $f$ ($Y^{\rm eq}_{f}$), appearing in Eq.~(\ref{be-s1}), are given respectively by~\cite{Gondolo:1990dk,Edsjo:1997bg}
\begin{equation}
\langle {\sigma v}_{f\bar{f}\rightarrow S_{1}^1 S_1^1}\rangle=\frac{z}{8 M_f^4 M_{\rm sc} {K_2(z M_f/M_{\rm sc})}^2}\int_{4 M_f^2}^{\infty}{{\sigma}_{f\bar{f}\rightarrow S_{1}^1 S_1^1}~\left(s-4 M_f^2\right)\sqrt{s}~K_1\left(\frac{z \sqrt{s}}{M_{\rm sc}}\right)ds}, 
\label{thermal-annihilation}
\end{equation}
\begin{eqnarray}
Y^{\rm eq}_{f} = \frac{45~g_{f}}{4 \pi^4} \left(\frac{z M_f}{M_{\rm sc}}\right)^2 \frac{ K_{2}\left(z M_f/M_{\rm sc}\right)}{g_{s}
(M_{\rm sc}/z)}\,,
\label{thermal-equilibrium}
\end{eqnarray}
where $g_f$ is the internal d.o.f of $f$, $K_{1}(z)$ and $K_{2}(z)$ are the Bessel function for first and second kind, respectively, and ${\sigma}_{f\bar{f}\rightarrow S_{1}^1 S_1^1}$ is given in Ref.~\cite{Iso:2010mv}. 
Solving the Boltzmann equation given by Eq.~(\ref{be-s1}) gives us the co-moving number
density $Y_{S_1^1}$. The corresponding 
relic density of the WDM $S_1^1$ can be calculated by using the following formula~\cite{Edsjo:1997bg},
\begin{eqnarray}
\Omega\, h^{2} \simeq 2.755 \times 10^{8}\, \left( \frac{M_{S_1^1}}{\rm GeV}\right)\,Y_{S_1^1}(\infty)\,.
\label{relic-density-expressionS1}
\end{eqnarray}
\begin{figure}[t!]
\centering
\includegraphics[angle=0,height=6.5cm,width=7.5cm]{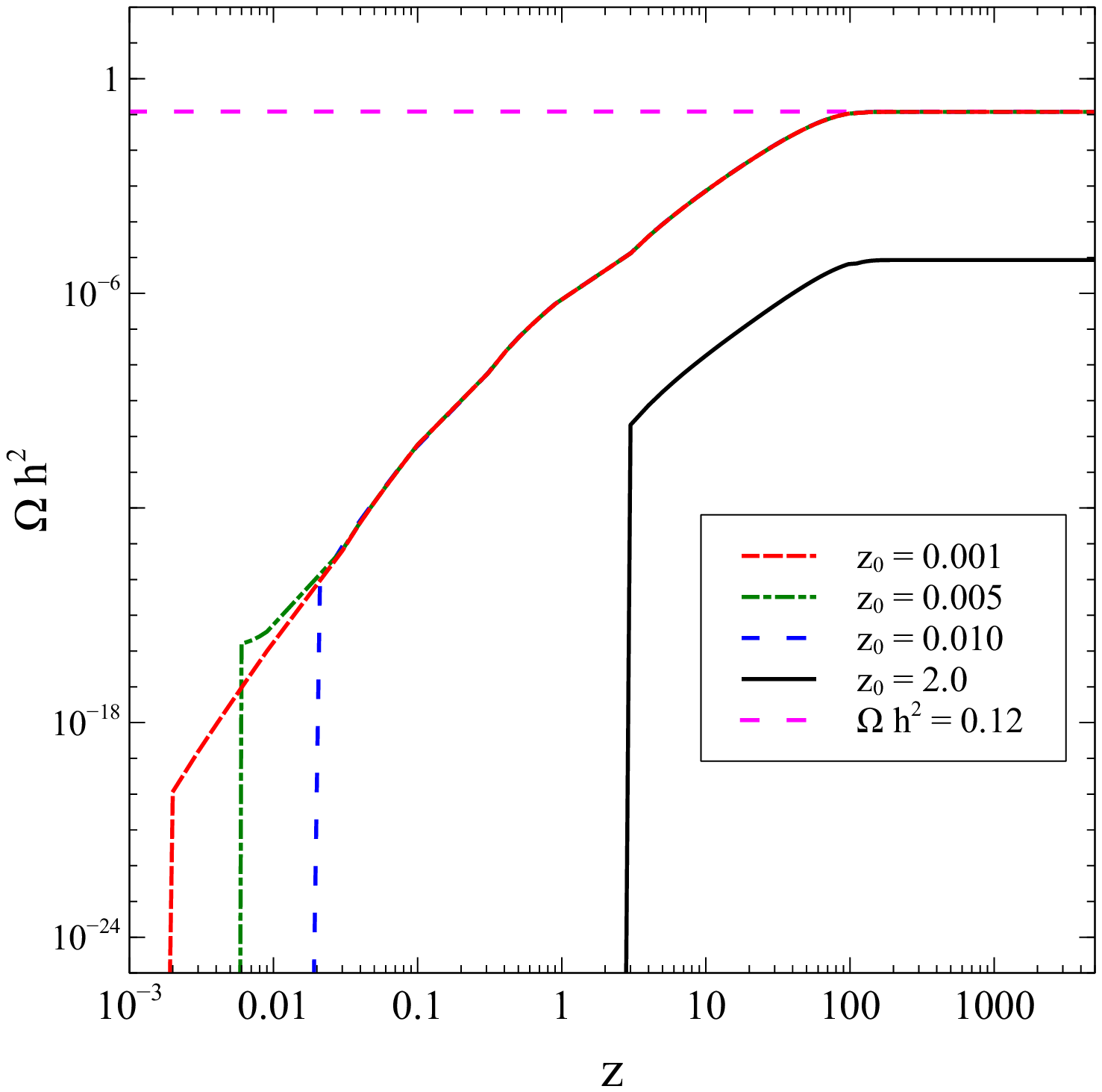}~~
\includegraphics[angle=0,height=6.5cm,width=7.5cm]{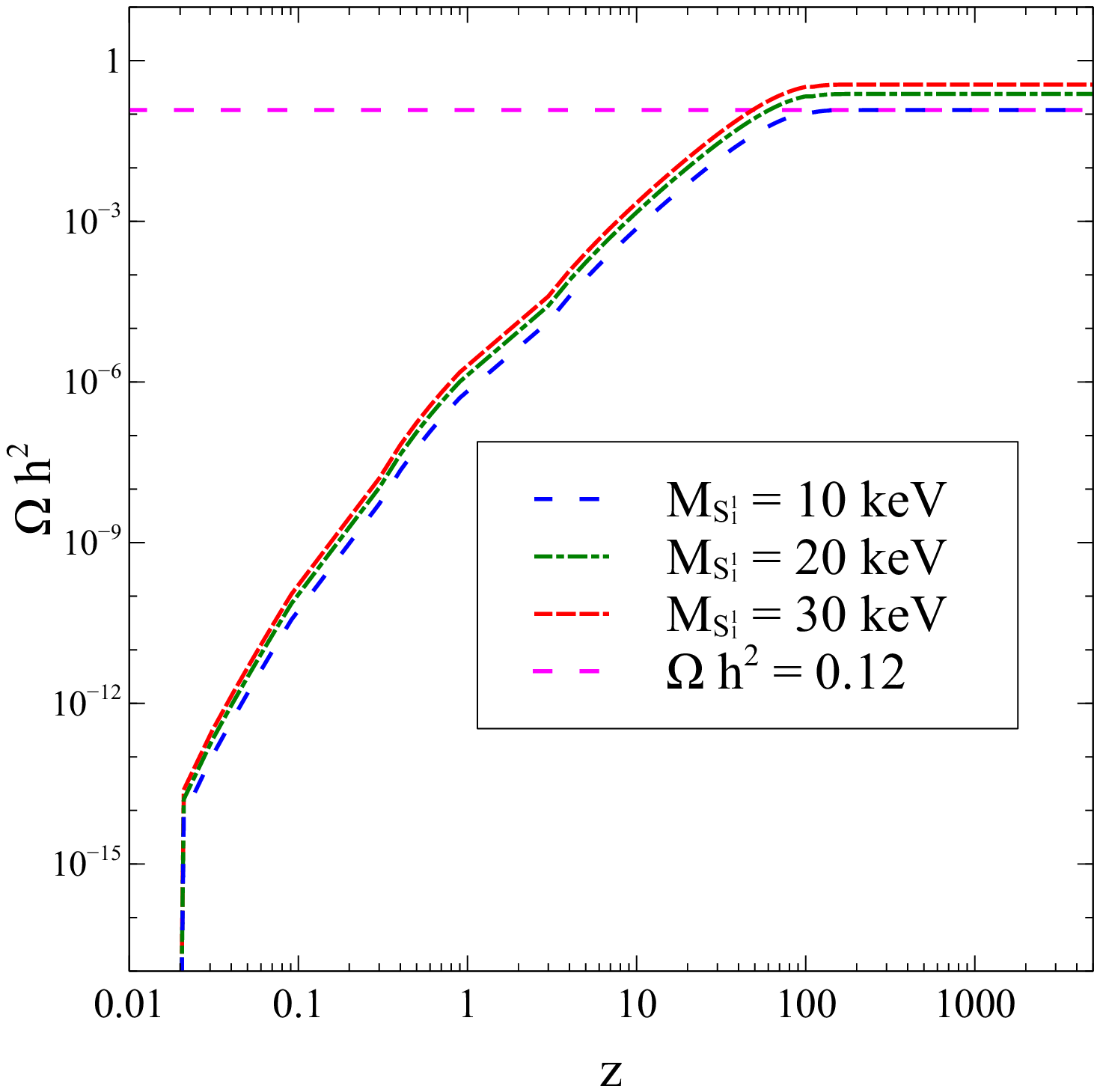}
\caption{Variation of DM relic
density, for the decay contribution only, as a function of $z$ for different values of $z_0$ (left), where $M_{S^1_1} = 10$~keV and different values of the WDM mass $M_{S_1^1}$ (right), where $z_0 = 0.01$. Here $g' = 12.5 \times 10^{-10}$, $M_{Z'} = 10$~GeV,
$M_{h_2} = 1$~TeV, and $\alpha = 0.01$~rad.}
\label{b}
\end{figure}
In order to understand the relative contribution of the decay and annihilation channels we will first consider them one at a time and solve the Boltzmann equation to get the relic density. We start with taking only the decay contribution and show in the 
left and right panels of Fig.~\ref{b} the variation of DM relic
density as a function of $z$, for different values of the initial temperature $T_0$ ($= M_{\rm sc}/z_0$) and different values of the WDM mass $M_{S_1^1}$, respectively. The horizontal magenta dashed line refers to the DM relic density measurement ($\Omega h^2\simeq 0.12$)~\cite{Ade:2015xua}. In the left panel,
as long as $T_0$ 
is greater than the mass of the mother particles ($h_{1,2}$) in $h_{1,2}\to Z'Z'$ decay channels, the final DM relic density is insensitive to $T_0$, as seen for the ${{z_0= 0.001,0.005,0.01}}$ cases. However, once
$T_0$ drops below the mass of the mother particles (presently
${{z_0 = 2}}$ case), $Z'$ 
production gets Boltzmann suppressed and consequently DM relic density is suppressed. In the right panel we show the dependence of the DM relic density on its 
mass ($M_{S_1^1}$) for $z_0=0.01$. It is clear that the relic density increases
with the DM mass, as expected from
Eq.~(\ref{relic-density-expressionS1}).
\begin{figure}[t!]
\centering
\includegraphics[angle=0,height=5.5cm,width=7.5cm]{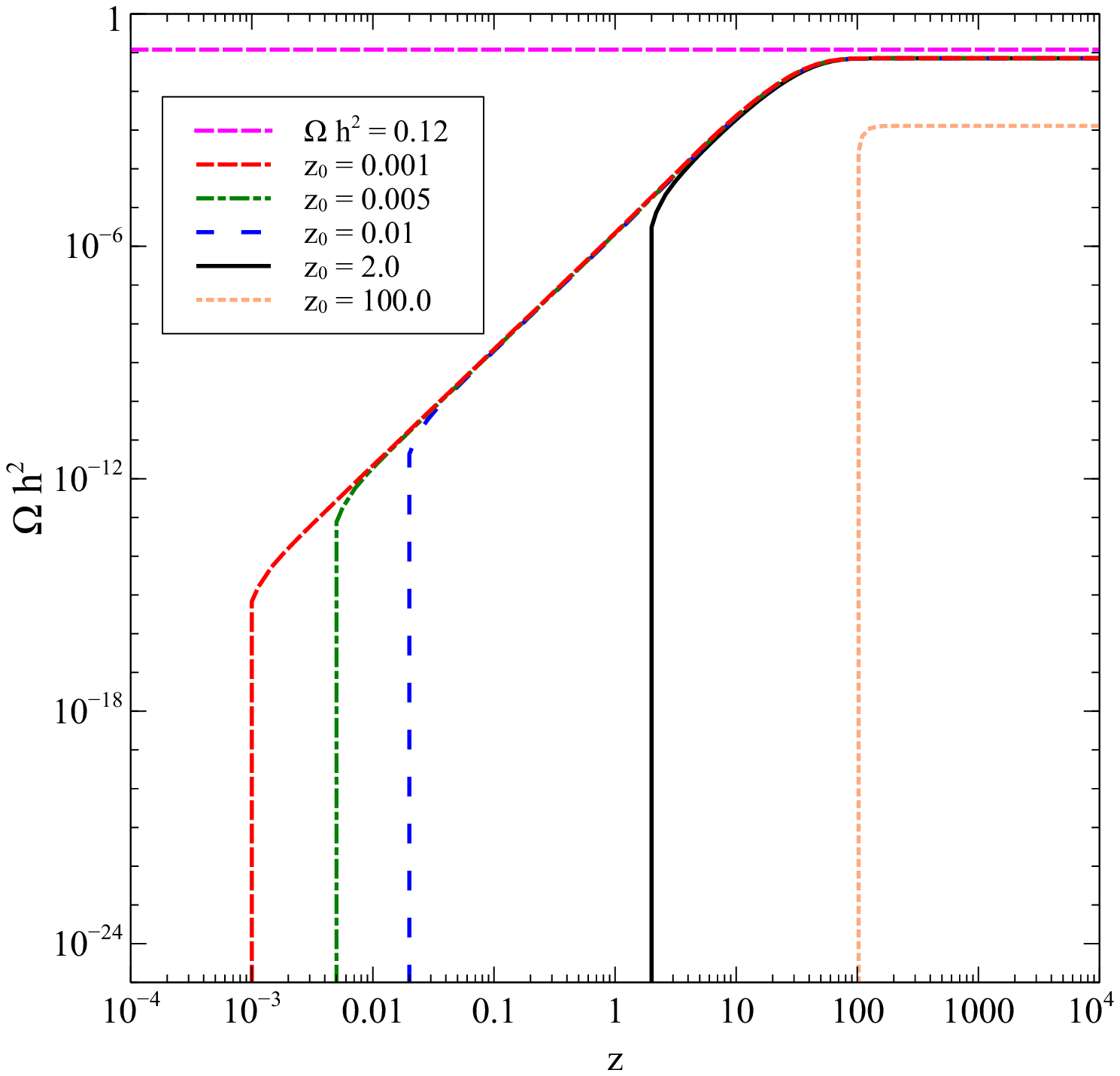}
\includegraphics[angle=0,height=5.5cm,width=7.5cm]{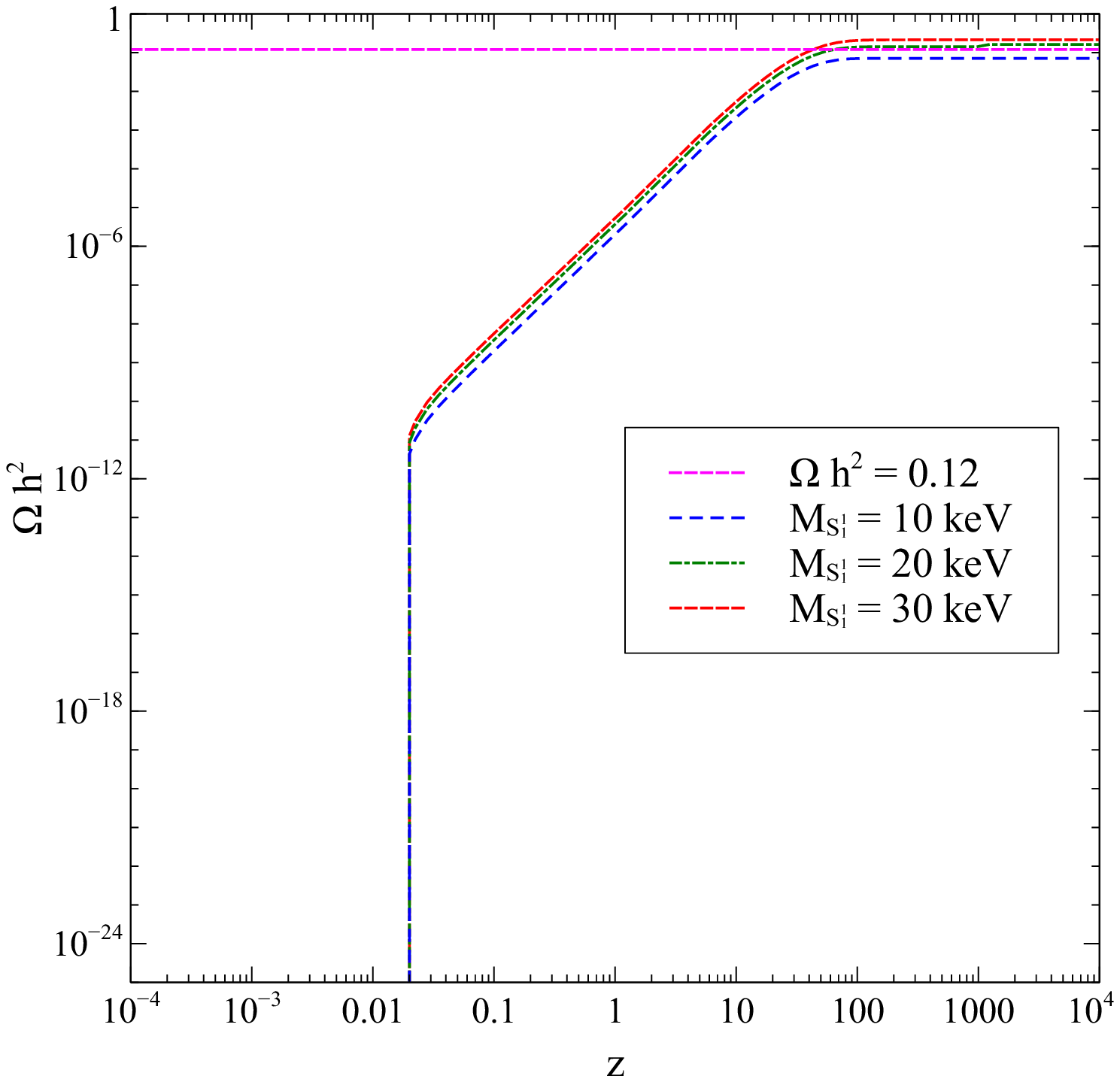}
\caption{Variation of DM relic
density, for the annihilation contribution only, as a function of $z$ for different values of $z_0$ (left), where $M_{S^1_1} = 10$~keV and different values of the WDM mass $M_{S_1^1}$ (right), where $z_0 = 0.01$. Here, $g' = 12.5 \times 10^{-10}$, $M_{Z'} = 10$~GeV,
$M_{h_2} = 1$~TeV, and $\alpha = 0.01$~rad.}
\label{bANN}
\end{figure}

For the annihilation contribution, $f\bar{f} \to S_{1}^1 S_1^1$,  there are two relevant  regimes are as follows. $(i)$ The on-shell regime, where $2 M_{S^1_1}< M_{Z'}< T_0$, in which  $Y_{S_1^1} (\propto g'^4/\Gamma_{Z'})$ does not depend on $T_0$ and $(ii)$ The  EFT regime,  where $2 M_{S^1_1}< T_0 < M_{Z'}$, in which  $Y_{S_1^1} (\propto g'^4 T^3_0/M_{Z'}^4)$ depends on $T_0$. In the left panel of Fig.~\ref{bANN} we see that 
as long as $T_0$ 
is greater than $M_{Z'}$, the final DM relic density is insensitive to $T_0$ (on-shell regime), as seen for the $z_0= 0.001,0.005,0.01,2.0$ cases. Once
$T_0$ drops below $M_{Z'}$ (presently
${{z_0 = 100.0}}$ case), then  $S_1^1$ production gets 
the suppressed by a factor $T_0^3/M_{Z'}^4$ (EFT regime). In the right panel of Fig.~\ref{bANN},  we show the dependence of the DM relic density on its mass ($M_{S_1^1}$) for $z_0=0.01$ (on-shell regime).  It is clear that the relic density increases
with the DM mass, as expected from
 Eq.~(\ref{relic-density-expressionS1}).

\begin{figure}[t!]
\centering
\includegraphics[angle=0,height=6.5cm,width=7.2cm]{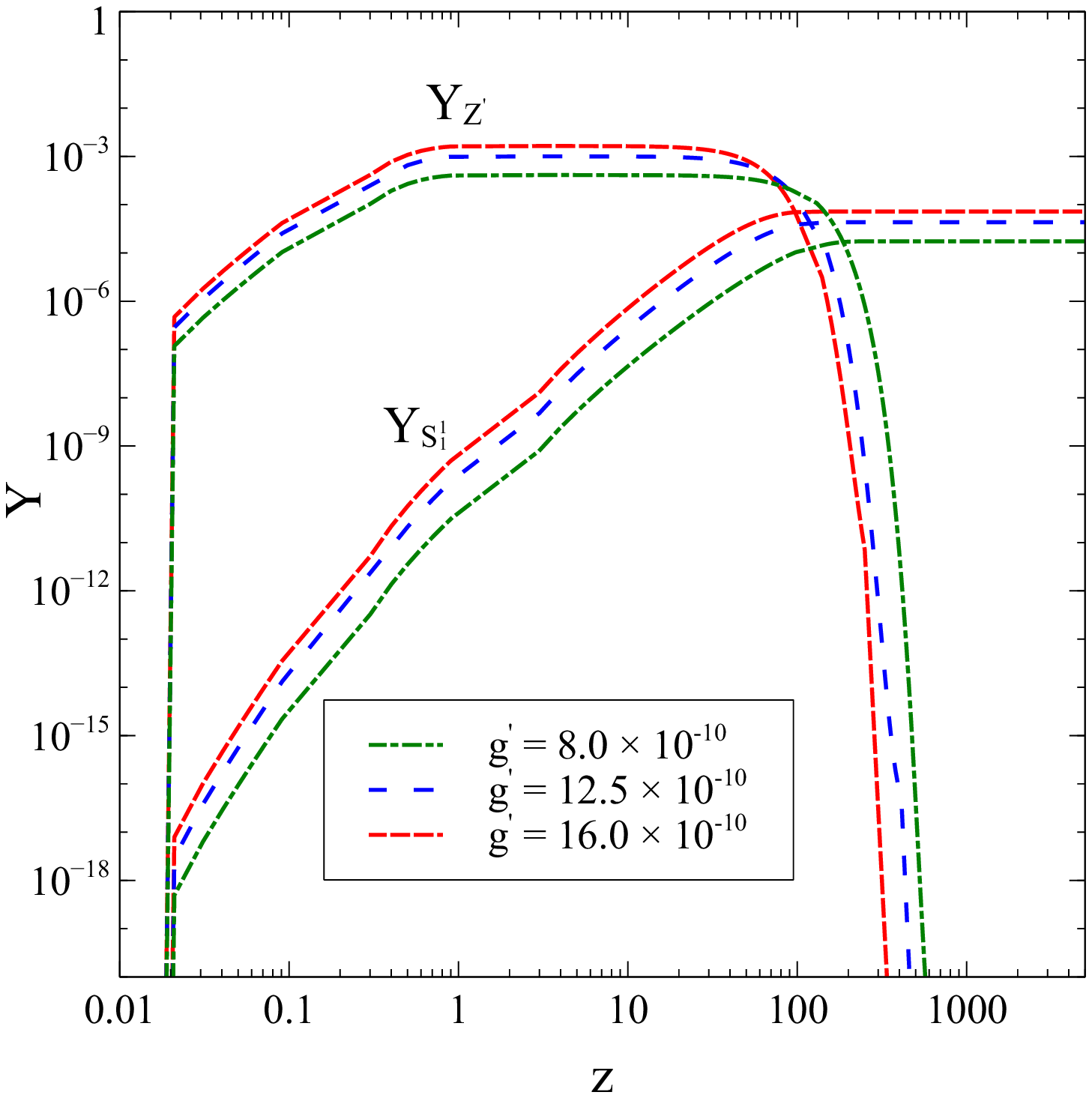}
\includegraphics[angle=0,height=7.70cm,width=8.0cm]{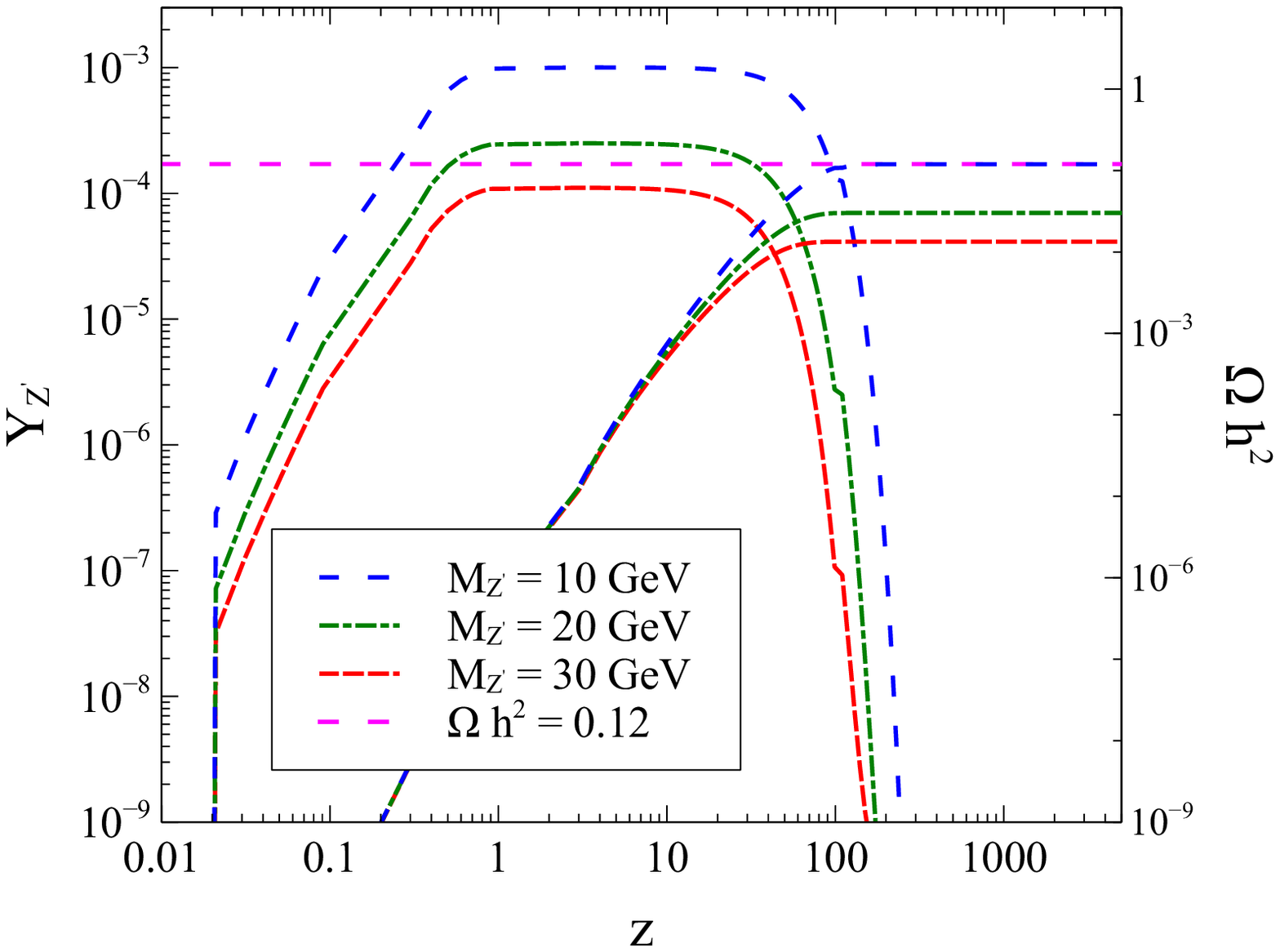}
\caption{Variation
of the co-moving number density of $Z'$
and the WDM $S_1^1$ (for the decay contribution) as a function of $z$ for different values of $B-L$
gauge coupling $g'$ (left) where $M_{Z'} = 10$~GeV and $M_{Z'}$ (right), where $g' = 12.5\times 10^{-10}$. Here, $z_0=0.01$, $M_{S_1^1} = 10$~keV,
$M_{h_2} = 1$~TeV, and $\alpha = 0.01$~rad.}
\label{c}
\end{figure}
\begin{figure}[t!]
\centering
\includegraphics[angle=0,height=6.2cm,width=7.5cm]{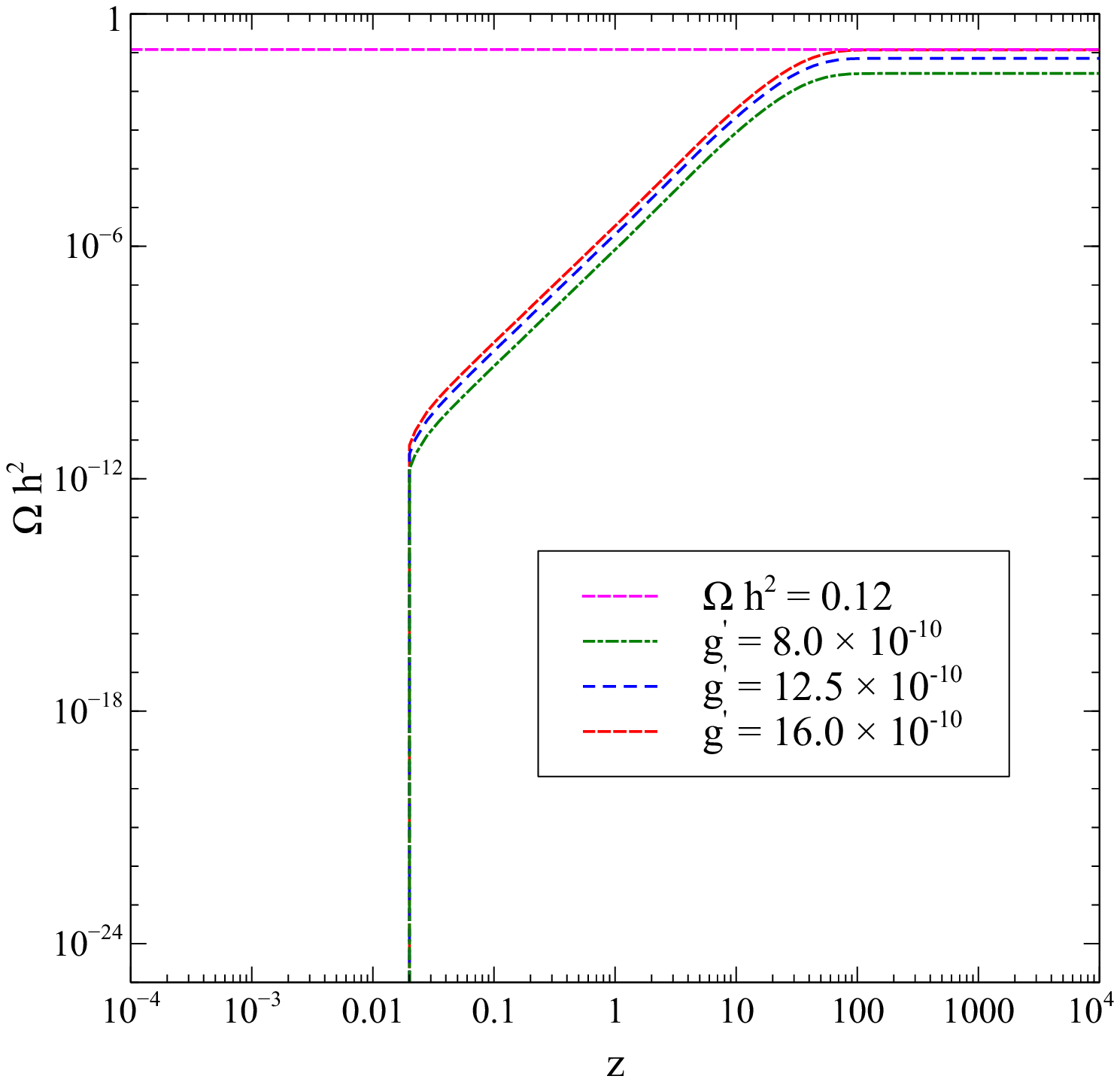}
\includegraphics[angle=0,height=6.2cm,width=7.5cm]{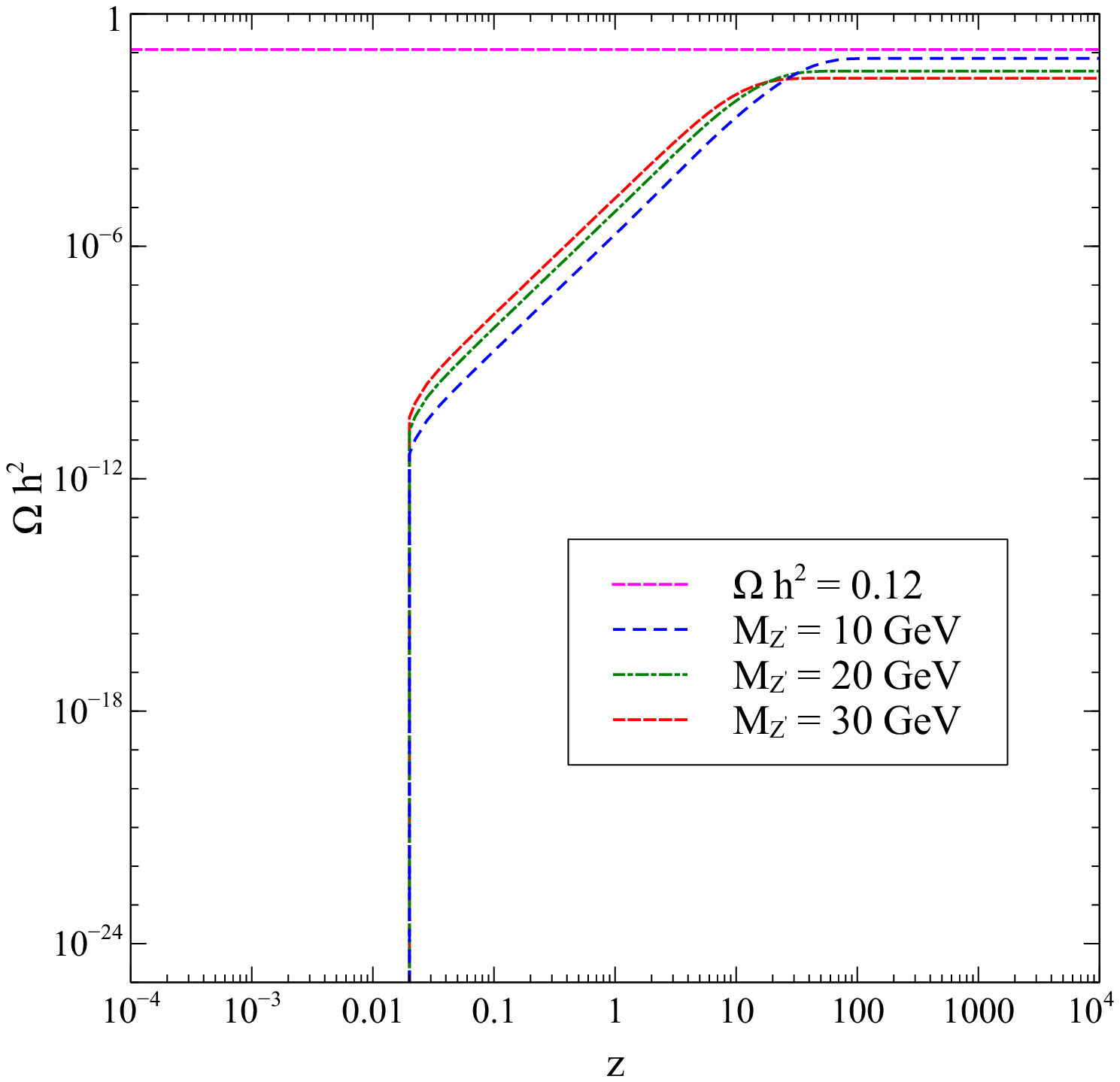}
\caption{Variation of DM relic
density, for the annihilation contribution, as a function of $z$ for different values of $B-L$
gauge coupling $g'$ (left) where $M_{Z'} = 10$~GeV and $M_{Z'}$ (right), where $g' = 12.5\times 10^{-10}$. Here, $z_0=0.01$, $M_{S_1^1} = 10$~keV,
$M_{h_2} = 1$~TeV, and $\alpha = 0.01$~rad.}
\label{cANN}
\end{figure}

For the decay contribution ($Z'\to S_1^1 S_1^1$), we show in Fig.~\ref{c} the variation
of the co-moving number density of $Z'$ (left panel)
and the co-moving number density of $S_1^1$ with $z$, for different values of $B-L$
gauge coupling $g'$ (left panel) and $M_{Z'}$ (right panel). Since $Z'$ production is proportional
to the $B-L$ gauge coupling $g'$, larger $g'$ results in larger $Z'$ production and consequently  a larger production of DM, as seen in the left panel of this
figure. Note also that $Z'\to S_1^1 S_1^1$ decay rate is directly proportional
to $g'$, and hence increasing $g'$ increases the decay rate of $Z'$ and hence the abundance of  $S_1^1$. Therefore, it is clear that for higher values
of $g'$ the $Z'$ co-moving number density plateau starts
bending at smaller values of $z$. On the other hand, in the right panel of Fig.~\ref{c}, we see that by increasing $M_{Z'}$ exactly opposite behavior appears
for $Z'$ production while similar
behavior happens for its decay. As mentioned, $Z'$ production mainly happens through decay of the Higgs bosons ($h_{1,2}$)
and those decay modes ($\Gamma_{h_i\to Z' Z'}$) are proportional to $M^3_{h_i}/M_{Z'}^2$ [see Eq.~(\ref{dec-expression-h_i-zz})].  Therefore, increasing $M_{Z'}$ reduces the production rate of $Z'$ as $1/M_{Z'}^2$. However,  its decay width is simply
proportional to its mass $M_{Z'}$ [see Eq.~(\ref{zpdecay})] and so increasing 
$M_{Z'}$ results in faster decay of $Z'$. In Fig.~\ref{cANN} we show similar plots for the annihilation contribution. This figure shows features similar  to Fig.~\ref{c}.

\begin{figure}[t!]
\centering
\includegraphics[angle=0,height=6.5cm,width=8.5cm]{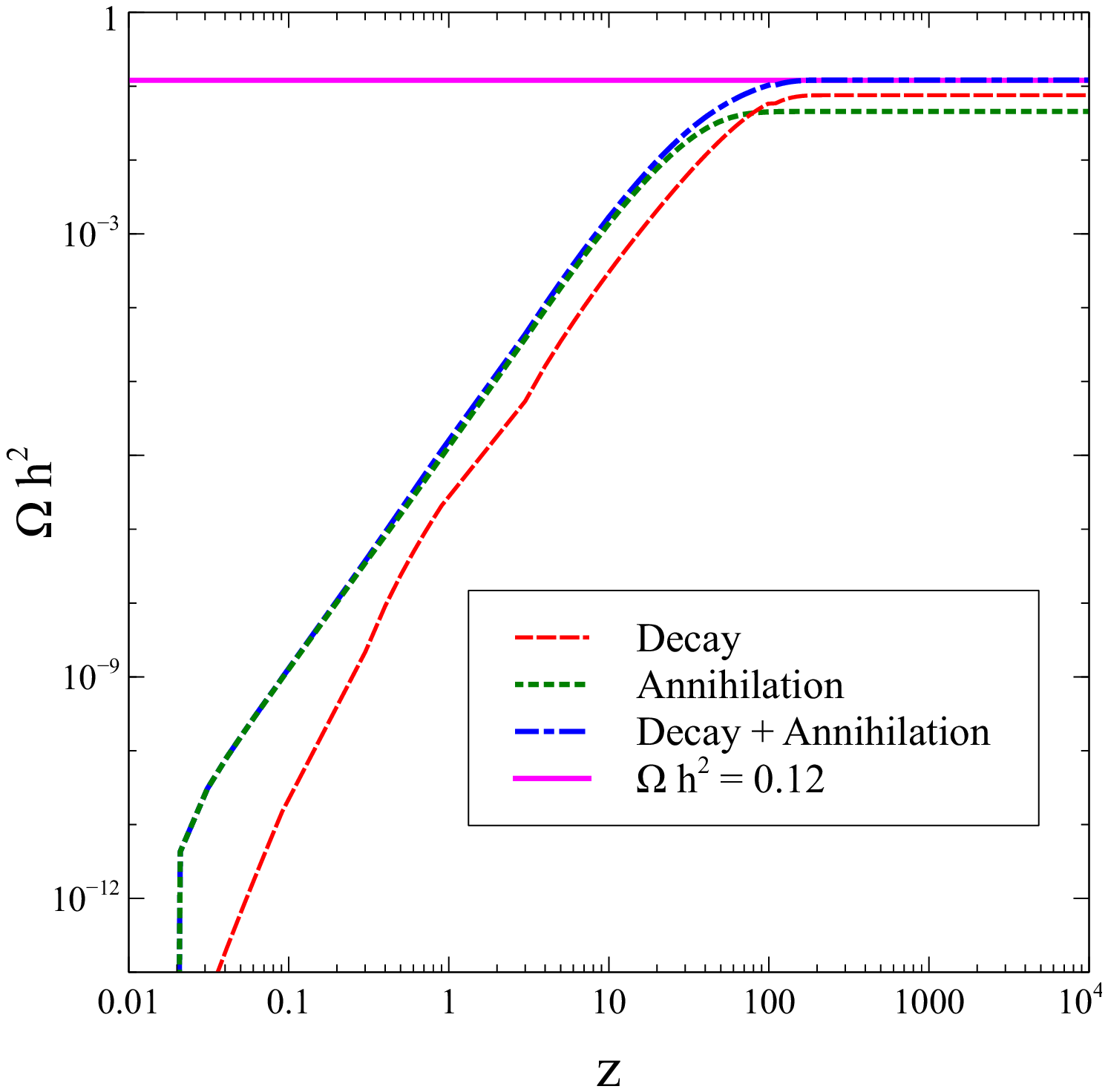}
\caption{Variation of relic density contributions of the single DM component scenario as a function of~$z$. Here,
 $M_{S_1^1} = 10$~keV, $M_{Z'} = 10$~GeV,
$g' = 1.0 \times 10^{-9}$, $M_{h_2}$ = 1~TeV, $z_0=0.01$, and $\alpha=0.01$~rad.}
\label{best-fit-plot}
\end{figure}
In Fig.~\ref{best-fit-plot}, we show the total relic density (blue dashed-dotted line) 
as well as the relative contributions of the two different types of WDM production processes, decay (red dashed line) and annihilation (green dotted line). Here, for a suitably selected set of model parameters ($M_{S_1^1} = 10$~keV, $M_{Z'} = 10$~GeV,
$g' = 1.0 \times 10^{-9}$, $M_{h_2}$ = 1~TeV, $z_0=0.01$, and $\alpha=0.01$~rad), the total WDM relic density equals the observed relic density ($\Omega h^2\simeq 0.12$) at the present epoch, where decay contributes $\sim 62\%$ of the WDM relic density while the rest comes from annihilation. It is worth mentioning that initially for $z \lesssim 100$, WDM is dominantly produced from the 
annihilation processes this is because of all ingoing particles
are already in the cosmic soup, while for $z \gtrsim 100$, the decay process starts
dominating, as seen in Fig.~\ref{c}.

\begin{figure}[t!]
\centering
\includegraphics[angle=0,height=5.5cm,width=7.5cm]{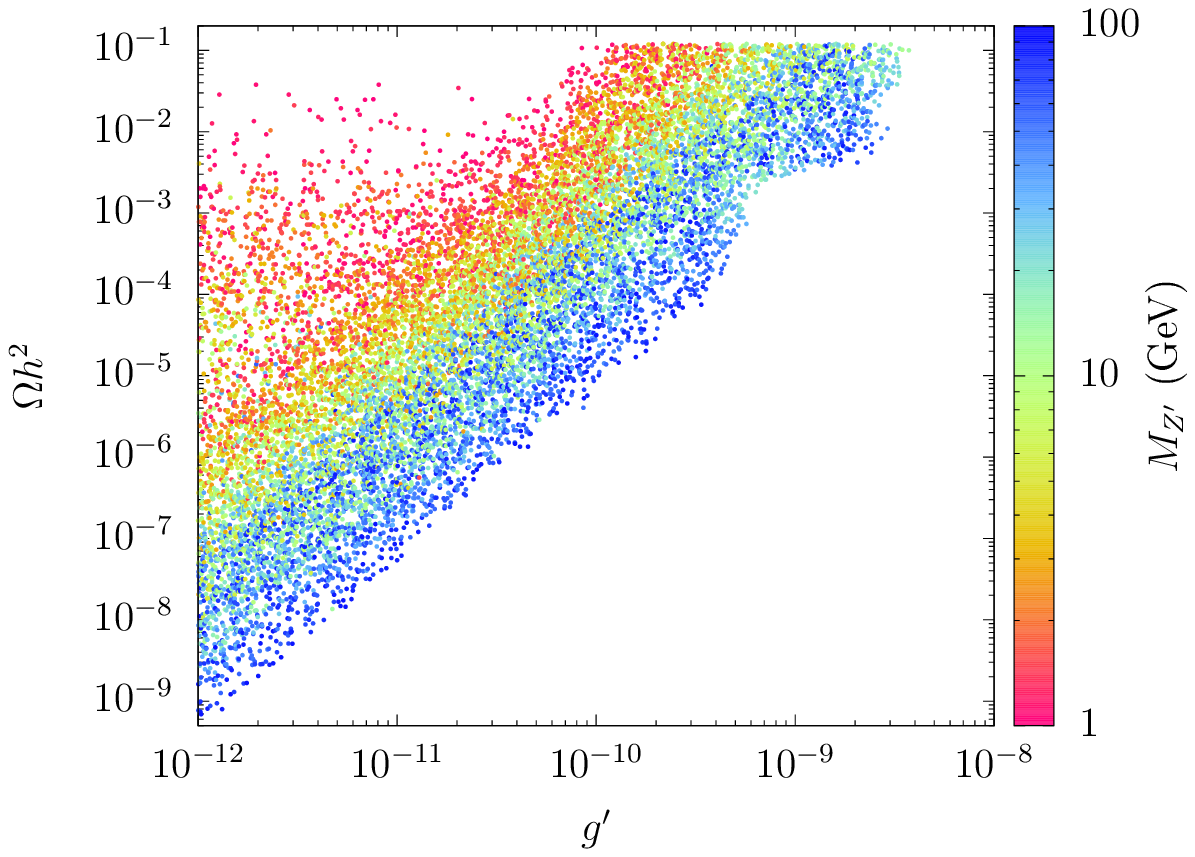}
\includegraphics[angle=0,height=5.5cm,width=7.5cm]{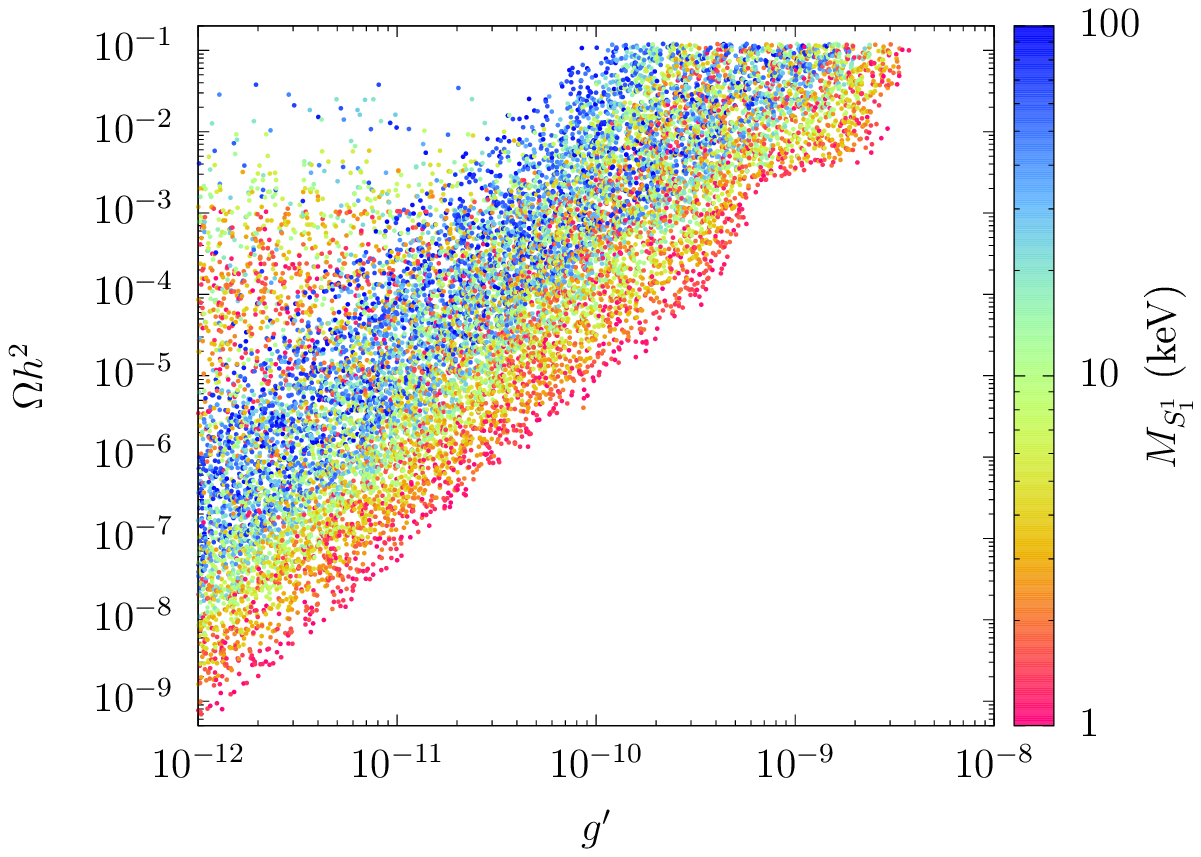}
\caption{Allowed points in $(g^{\prime},\Omega h^2)$
plane after imposing a constraint $\Omega h^2 \leq 0.12$, as an upper bound on the WDM relic density $\Omega h^2$.
The color bars of left and right panels  correspond to the $Z^{\prime}$ mass
($M_{Z'}$) in GeV and WDM mass ($M_{S_1^1}$) in keV, respectively.}
\label{scatt-1}
\end{figure}

Variation of total WDM relic density ($\Omega h^{2}$) as a function of the gauge coupling $g^{\prime}$ can be seen in Fig.~\ref{scatt-1},
where the BLSMIS points have been generated over the following ranges of its fundamental parameters: $1\leq M_{S_1^1} \leq 10$~keV, $1\leq M_{Z'} \leq 100$~GeV,
$10^{-12}\leq g' \leq 10^{-8}$, $M_{h_2} = 1$~TeV, $z_0=0.01$, and $\alpha=0.01$~rad. From the left panel it is clear that $\Omega h^{2}$ is inversely proportional to $M_{Z'}$ (which is represented by the color bar).
More explicitly, for a fixed $g'$ value,  larger $\Omega h^{2}$ values correspond to smaller $M_{Z'}$ values (red points)
and vice versa for the blue points. On the other hand as illustrated in the right panel, $\Omega h^{2}$ is directly proportional to the WDM mass $M_{S^1_1}$ (which is represented by the color bar). This is consistent with $\Omega h^{2}$ expression given in Eq.~(\ref{relic-density-expressionS1}). 
\begin{figure}[t!]
\centering
\includegraphics[angle=0,height=5.5cm,width=7.5cm]{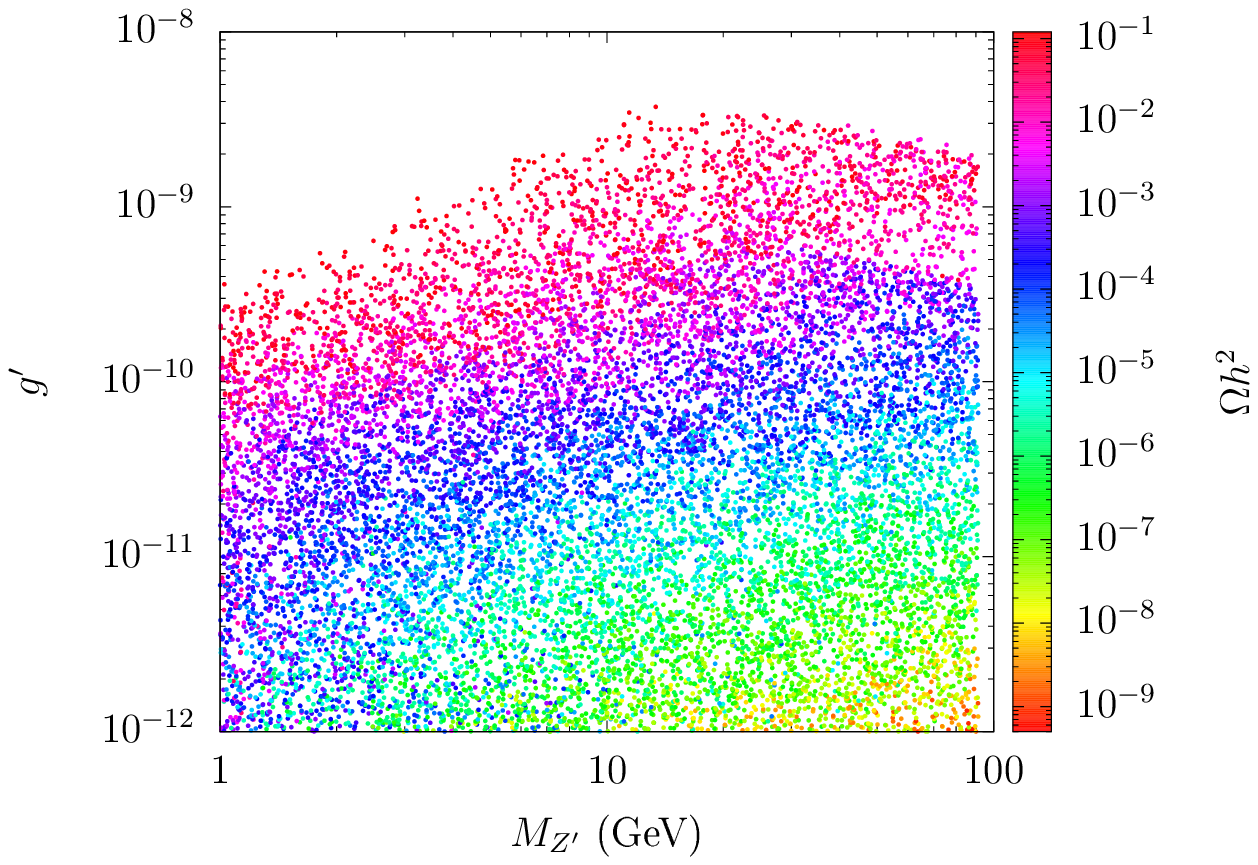}
\includegraphics[angle=0,height=5.5cm,width=7.5cm]{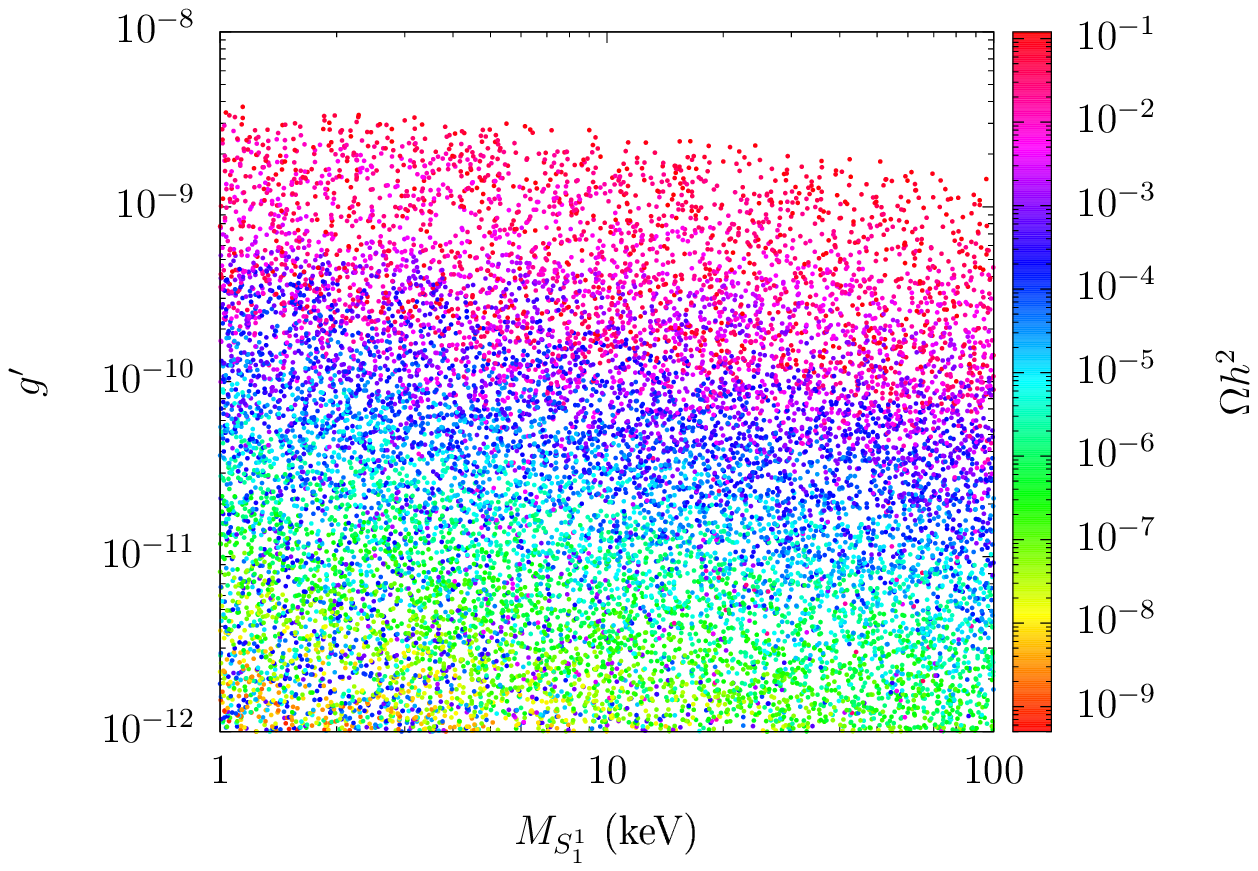}
\caption{Allowed points in $(M_{Z'},g')$ and $(M_{S_1^1}, g')$
planes after imposing an upper bound on the WDM relic density $\Omega h^2 \leq 0.12$. The color bars of left and right panels correspond to $\Omega h^{2}$.}
\label{scatt-2}
\end{figure} 

In Fig.~\ref{scatt-2} we show the allowed points
in the $(M_{Z'},g')$ and $(M_{S_1^1}, g')$ planes in the left and right panels, respectively, which give the relic density consistent with a relic density upper bound of the Planck measurement ($\Omega h^2 \leq 0.12$)~\cite{Ade:2015xua}. All other parameter values are allowed to vary in the range mentioned in the previous paragraph. From the figure color bars (mapped to the total WDM relic density $\Omega h^2$), 
it is clearly seen that many points ($\sim 84\%$~of the scanned points) have  a small DM relic density ($\Omega h^2\leq 10^{-2}$). Therefore, in the next section we discuss a two component FIMP DM possibility as a well-motivated scenario to get an extra relic density contribution from the lightest heavy RH neutrino, $\nu_H^1$, as a GeV scale DM. 
\section{Two component FIMP dark matter}\label{2DM}
In the previous section we have studied the WDM FIMP $S_1^1$, as a single component DM. As mentioned in section~\ref{BLSMIS}, the lightest heavy RH neutrino, $\nu_H^1$, can be long-lived particle by making the corresponding Yukawa couplings very small $\lesssim 3\times 10^{-26} ({\rm GeV}/M_N)^{1/2}$ \cite{Fiorentin:2016avj,DiBari:2016guw}. Therefore, it can be an additional DM component of mass of order GeV. Note that any interaction between $S_1^1$ and $\nu_H^1$ is completely forbidden. Thus in the present section, we consider a two component DM scenario with two DM candidates: the WDM FIMP $S_1^1$ and the lightest heavy RH neutrino $\nu_H^1$. The dominant annihilation channels of $\nu^1_H$ pair to SM particles 
 are mediated by $Z'$
and $h_i$ ($i=1,2$).\footnote{Due to the smallness of the corresponding Yukawa coupling of $\nu_H^1$ (as assumed to be a stable DM candidate), the contribution of the channels mediated by $Z$ and $W^\pm$ bosons is negligible. Also, the annihilation channels mediated by the SM-like Higgs $h_1$ are suppressed as compared to the $h_2$ ones, because the coupling $\lambda_{\nu^1_H \nu^1_H h_1}$ is very small since it is proportional to $\sin\alpha$ which is constrained to be very small by LHC \cite{Khachatryan:2016vau}.} The coupling strength of $\nu^1_H$ pair with $Z'$ is given by $g'/2$, while with $h_i$ ($i=1,2$) is given as
\begin{eqnarray}\label{HRHRH-coup}
\lambda_{\nu^1_H \nu^1_H h_i} 
= \sqrt{2} \ g' \ \frac{ M_{\nu^1_H} }{M_{Z'}}O_i\,,
\label{n1-n1-h-coupling}
\end{eqnarray}  
where $O_1=\sin\alpha$ and $O_2=\cos\alpha$. Therefore, $\nu_H^1$ pair annihilation is proportional
to the gauge coupling $g'$ which is taken very small in the present model. Due to this feeble gauge coupling $g'$, $\nu_H^1$ will never reach thermal equilibrium and is  produced by the freeze-in mechanism. The Boltzmann equation associated with $\nu_H^1$ production is as follows
\begin{eqnarray}
\frac{d Y_{\nu^1_H}}{dz} &=&  
\frac{2\, M_{\rm Pl}\, z\, \sqrt{g_{\star}}}{1.66\, M^2_{\rm sc}\, g_s}\,
\left[\langle \Gamma_{Z' \rightarrow \nu^1_H \nu^1_H} \rangle_{\rm NTH} 
\left(Y_{Z'} -  Y_{\nu^1_H} \right) +
\sum_{i=1,2}
\langle \Gamma_{h_i} \rangle \left(Y^{\rm eq}_{h_i} -  Y_{\nu^1_H} \right)
\right]\nn\\
&+&\frac{4 \pi^2}{45} \frac{M_{\rm Pl}\, M_{\rm sc} \sqrt{g_{\star}}}{1.66\, z^2}
\sum_{f}\langle {\sigma v}_{f\bar{f}\rightarrow \nu^1_H \nu^1_H} \rangle
\left[\left({Y_{f}^{\rm eq}}\right)^2 -  Y^2_{\nu^1_H} \right],
\label{be-N1}
\end{eqnarray}
where 
$ \langle \Gamma_{Z' \rightarrow \nu^1_H \nu^1_H} \rangle_{\rm NTH} $ and $\langle {\sigma v}_{f\bar{f}\rightarrow \nu^1_H \nu^1_H} \rangle$
are defined as in
Eqs.~(\ref{non-thermal-decay}),(\ref{thermal-annihilation}), respectively, by replacing $S_1^1$ with $\nu_H^1$, while $Y^{\rm eq}_{h_i}$ is defined as in Eq.~(\ref{thermal-equilibrium}) by replacing $f$ with $h_i$. Thermal average of the decay width of $h_i$ ($i=1,2$)
is defined as \cite{Gondolo:1990dk}
\begin{eqnarray}
\langle \Gamma_{h_i} \rangle = \frac{K_{1}(z)}{K_{2}(z)} \Gamma_{h_i}\,,
\end{eqnarray} 
where $\Gamma_{h_i}$ is the total decay width of $h_i$. After solving the Boltzmann equation of $\nu_H^1$ production,  Eq.~(\ref{be-N1}), the corresponding relic density
of $\nu^1_H$ can be determined by using the following relation,
\begin{eqnarray}
\Omega_{\nu^1_H} h^{2} = 2.755 \times 10^{8}\,
\left( \frac{M_{\nu^1_H}}{\rm GeV}\right)\,Y_{\nu^1_H}(\infty)\,.
\label{relic-density-expression}
\end{eqnarray}
Finally, the total
relic density of this two component DM  scenario is given by
\begin{eqnarray}
\Omega^{\rm tot} h^2 = \Omega_{\nu^1_H} h^{2} + \Omega_{S_1^1} h^{2}\,,
\end{eqnarray}
where $\Omega_{S_1^1} h^{2}$ is the relic density of 
$S_1^1$ which is defined in Eq.~(\ref{relic-density-expressionS1}).

It is clear that the DM production depends crucially on the DM mass  and the mass of the mother particles ($M_{Z'},M_{h_2}$). Assuming $M_{h_2}> 2 M_{Z'} > 4 M_{S_1^1}$, we divide the $\nu_H^1$ DM spectrum into two regions according to the dominant production modes of $\nu_H^1$ DM - Region~I, where $M_{Z'} > 2 M_{\nu_H^1}$ and $\nu_H^1$ production is $Z'$ dominated, and Region~II, where $M_{Z'} < 2 M_{\nu_H^1}$ and $\nu_H^1$ production is $h_2$ dominated. 
\subsection{\bf Region I: $M_{Z'} > 2 M_{\nu_H^1}$}
\label{mz>2mn_case}
\begin{figure}[t!]
\centering
\includegraphics[angle=0,height=6.5cm,width=8.5cm]{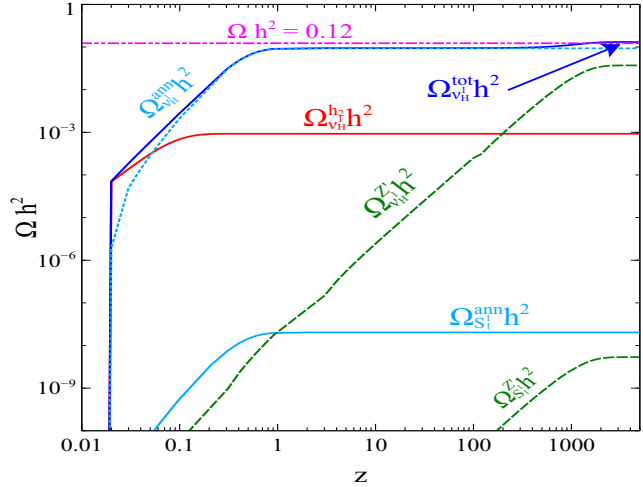}
\caption{Variation of relic density contributions of the two DM component scenario
as a function of~$z$. Here,
$M_{Z'} = 1$~TeV, $M_{\nu_H^1} = 70$~GeV, $M_{S_1^1} = 10$~keV, 
$g' = 9.0 \times 10^{-12}$, $M_{h_2}$ = 5~TeV, $\alpha=0.01$~rad, and~$z_0=0.01$.}
\label{fig-z-h}
\end{figure}
For our chosen set of BLSMIS parameters ($M_{Z'} = 1$~TeV, $M_{\nu_H^1} = 70$~GeV, $M_{S_1^1} = 10$~keV, 
$g' = 9.0 \times 10^{-12}$, $M_{h_2}$ = 5~TeV, $\alpha=0.01$~rad, and $z_0=0.01$), we show in Fig.~\ref{fig-z-h} the variation of the total DM relic density (blue solid line) and its relative contributions 
for a two component DM scenario. In the figure, red solid and green dashed lines correspond to the 
$\nu_H^1$ relic density contributions from the decay of $h_2$ and $Z'$, respectively\footnote{Due to a smallness of the mixing angle $\alpha$, DM production of the SM-like Higgs $h_1$ is negligible.}, while
cyan dashed line corresponds to the annihilation contribution ($\Omega^{\rm ann}_{\nu_H^1}h^2$). 
In addition, the 
$S_1^1$ relic density contribution from the decay of $Z'$ ($\Omega^{Z'}_{S_1^1}h^2$) and annihilation ($\Omega^{\rm ann}_{S_1^1}h^2$) are presented by green dashed and cyan solid lines, respectively. Note that in region~I, the relative contribution of $Z'$ decay to $\nu_H^1$ production ($\Omega^{Z'}_{\nu_H^1}h^2$) is larger than the $h_2$ decay contribution ($\Omega^{h_2}_{\nu_H^1}h^2$) because
the latter is suppressed by a factor of their partial decays ratio ($\Gamma_{h_2\to \nu_H^1 \nu_H^1}/\Gamma_{Z'\to \nu_H^1 \nu_H^1}$
$\simeq 12 M^2_{\nu_H^1}M_{h_2}/M^3_{Z'}\simeq {\cal O}(0.1)$). It is also worth noting that the relic density contribution of the keV DM ($S_1^1$) is negligible compared to the GeV DM ($\Omega_{\nu_H^1}h^2$) even though  
they have the same gauge coupling strength~($g'$) and their mediator masses ($M_{h_2}$ and $M_{Z'}$) are of the same order ($\sim$~TeV). This is simply because the relic density of a DM candidate is
directly proportional to its mass [see Eqs.~(\ref{relic-density-expressionS1}),(\ref{relic-density-expression})]. Therefore, the contribution of the keV mass $S_1^1$ to the DM total relic density is suppressed by a factor  $\simeq M_{S_1^1}/M_{\nu_H^1}\simeq {\cal O}(10^{-7})$ as compared to the GeV mass $\nu_H^1$.

\begin{figure}[t!]
\centering
\includegraphics[angle=0,height=6.3cm,width=7.5cm]{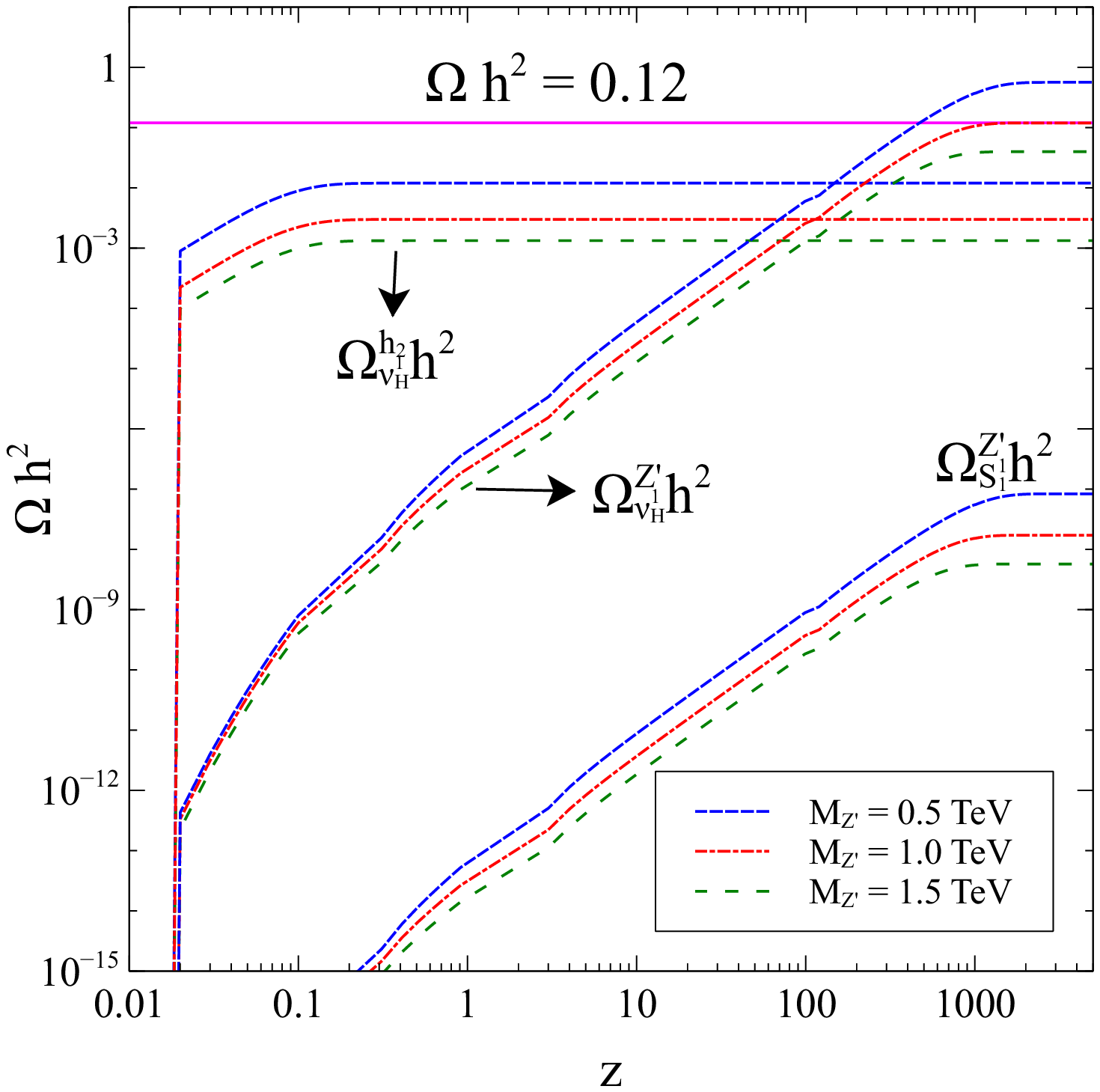}~~
\includegraphics[angle=0,height=6.3cm,width=7.5cm]{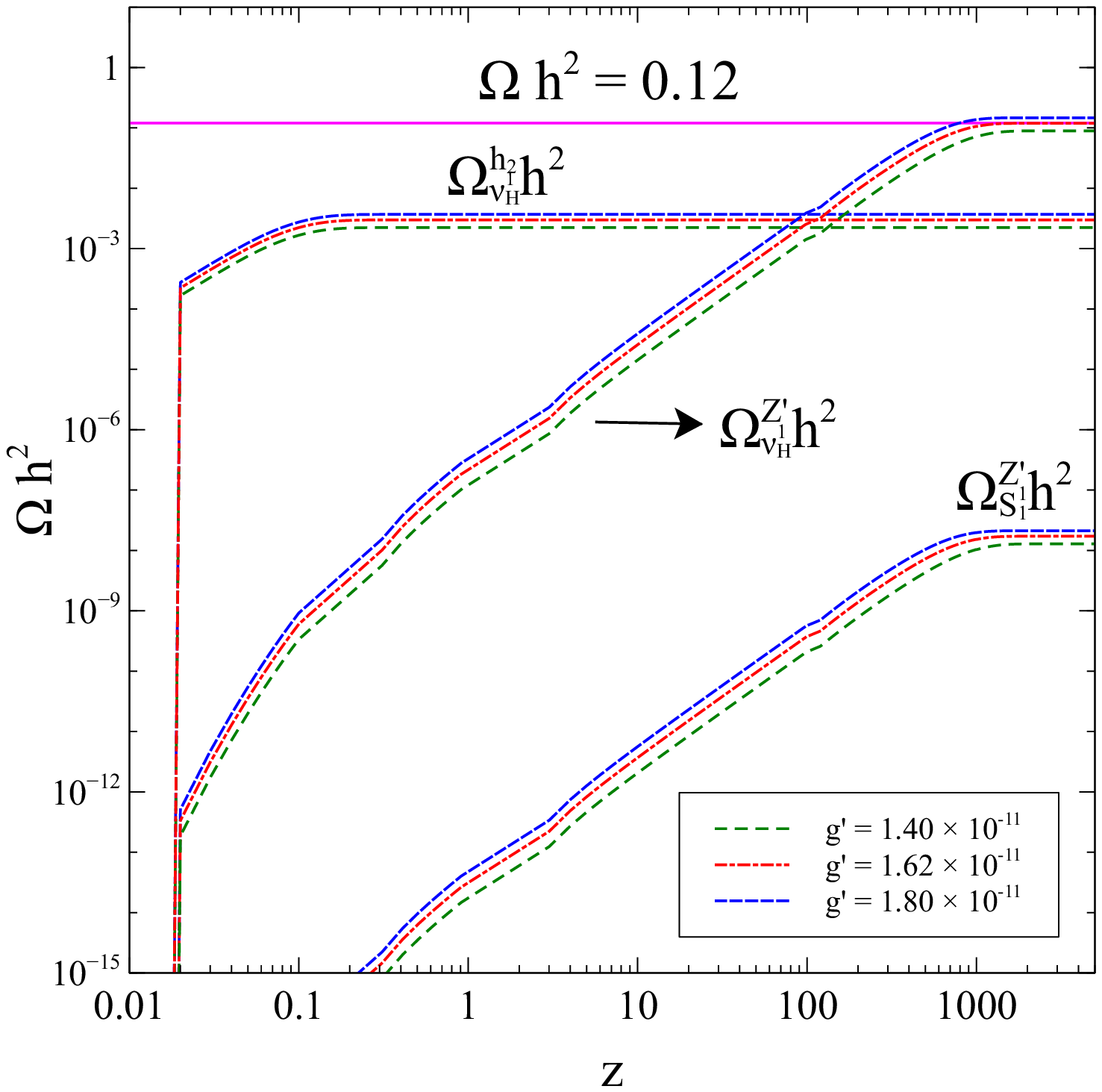}\\[0.5cm]
\includegraphics[angle=0,height=6.3cm,width=7.5cm]{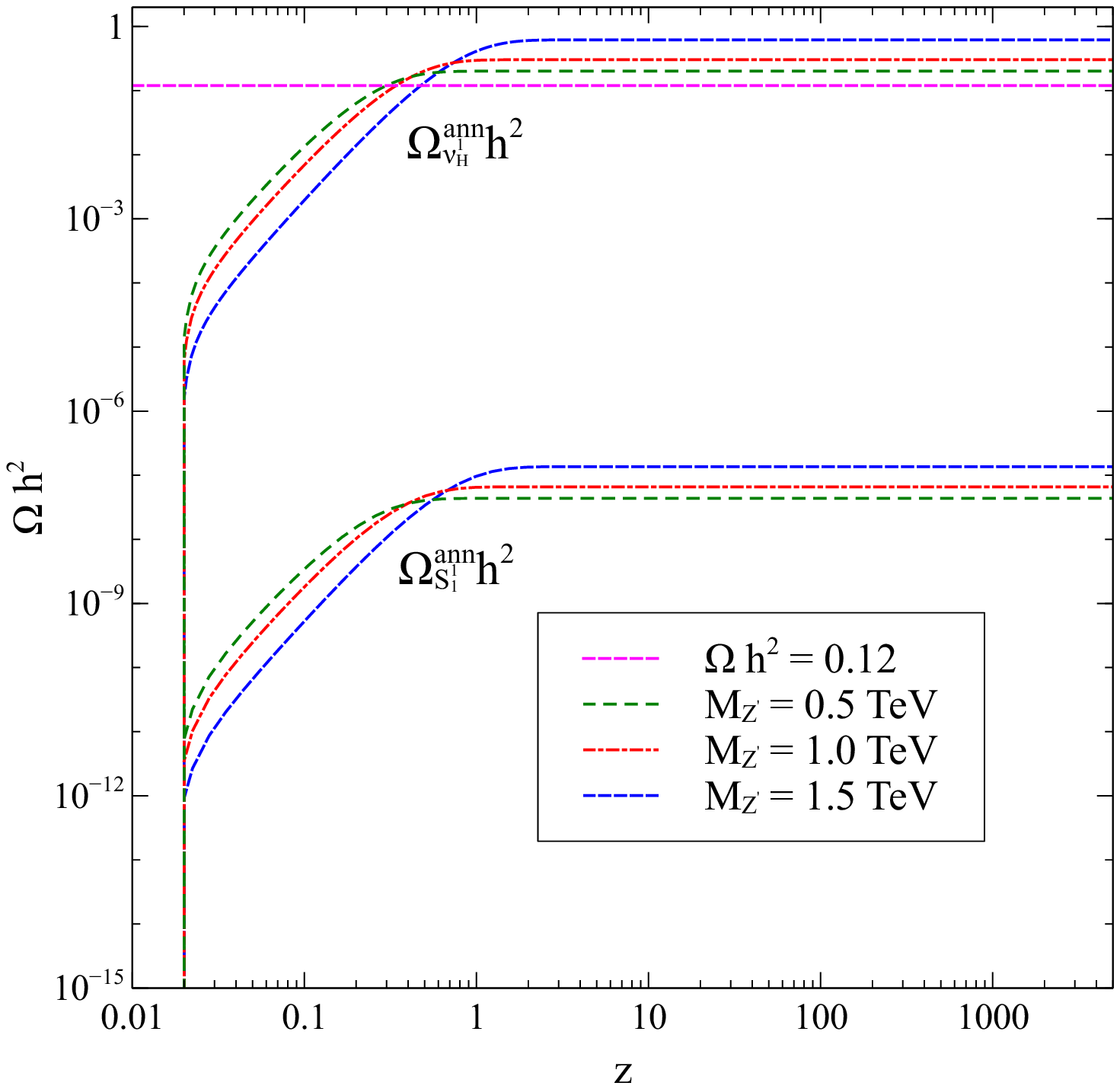}~~
\includegraphics[angle=0,height=6.3cm,width=7.5cm]{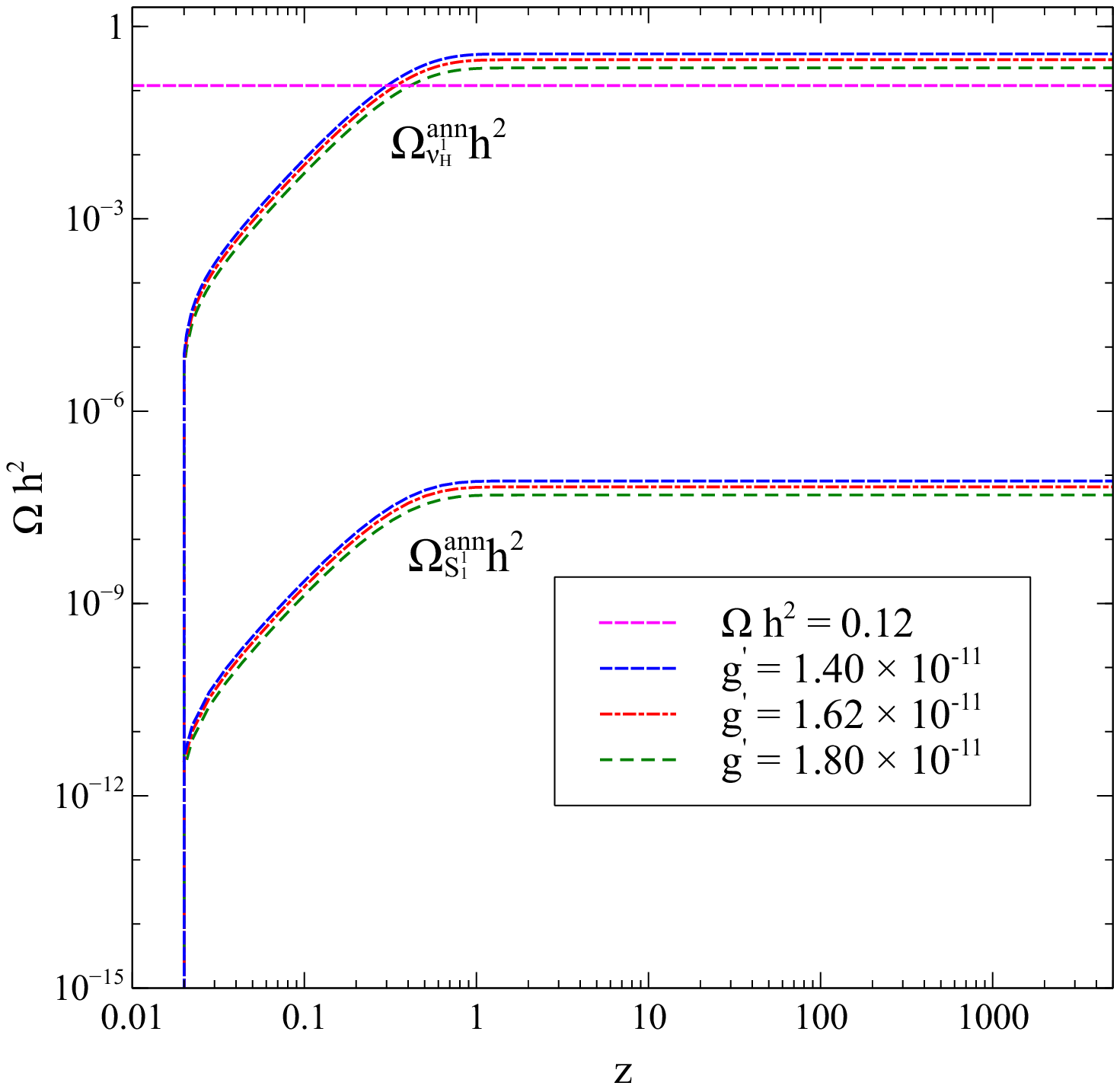}
\caption{Variation of relative relic density contributions of a two DM component scenario as a function of $z$ for different values of $M_{Z'}$ (left panels) and $g'$ (right panels). Top figures correspond to decay contributions while the bottom ones correspond to annihilation contributions. Here, $M_{Z'} = 1$~TeV, $M_{\nu_H^1} = 70$~GeV, $M_{S_1^1} = 10$~keV,
$g' = 1.62 \times 10^{-11}$, $M_{h_2}$ = 5~TeV, $\alpha=0.01$~rad, and $z_0=0.01$.}
\label{fig-z-h-a}
\end{figure}
In the left and right panels of Fig.~\ref{fig-z-h-a}, we show the variation
of relic density contributions of the two component DM scenario for different values of
$M_{Z'}$ and $g'$, respectively. The top panels stand for the decay contribution while the bottom ones stand for the annihilation contribution. Again from these figures, one can easily conclude that FIMP relic density contributions are inversely proportional to the mediator mass, as illustrated in left panels, and directly proportional to coupling strength as shown in right panels. We have discussed these features before in section~\ref{WDMFIMP}.    
\begin{figure}[t!]
\centering
\includegraphics[angle=0,height=5.5cm,width=7.5cm]{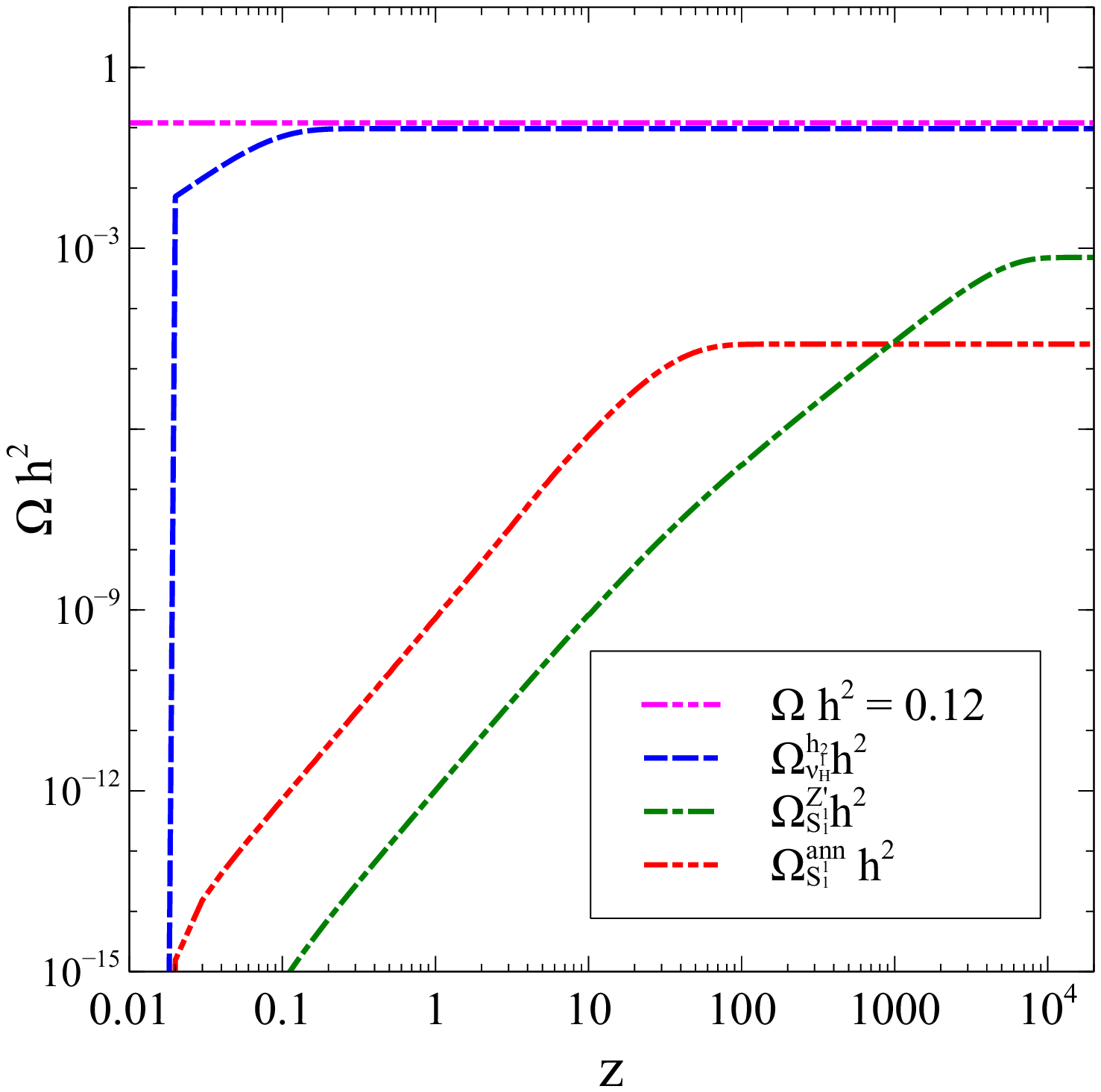}~~~\includegraphics[angle=0,height=5.5cm,width=7.5cm]{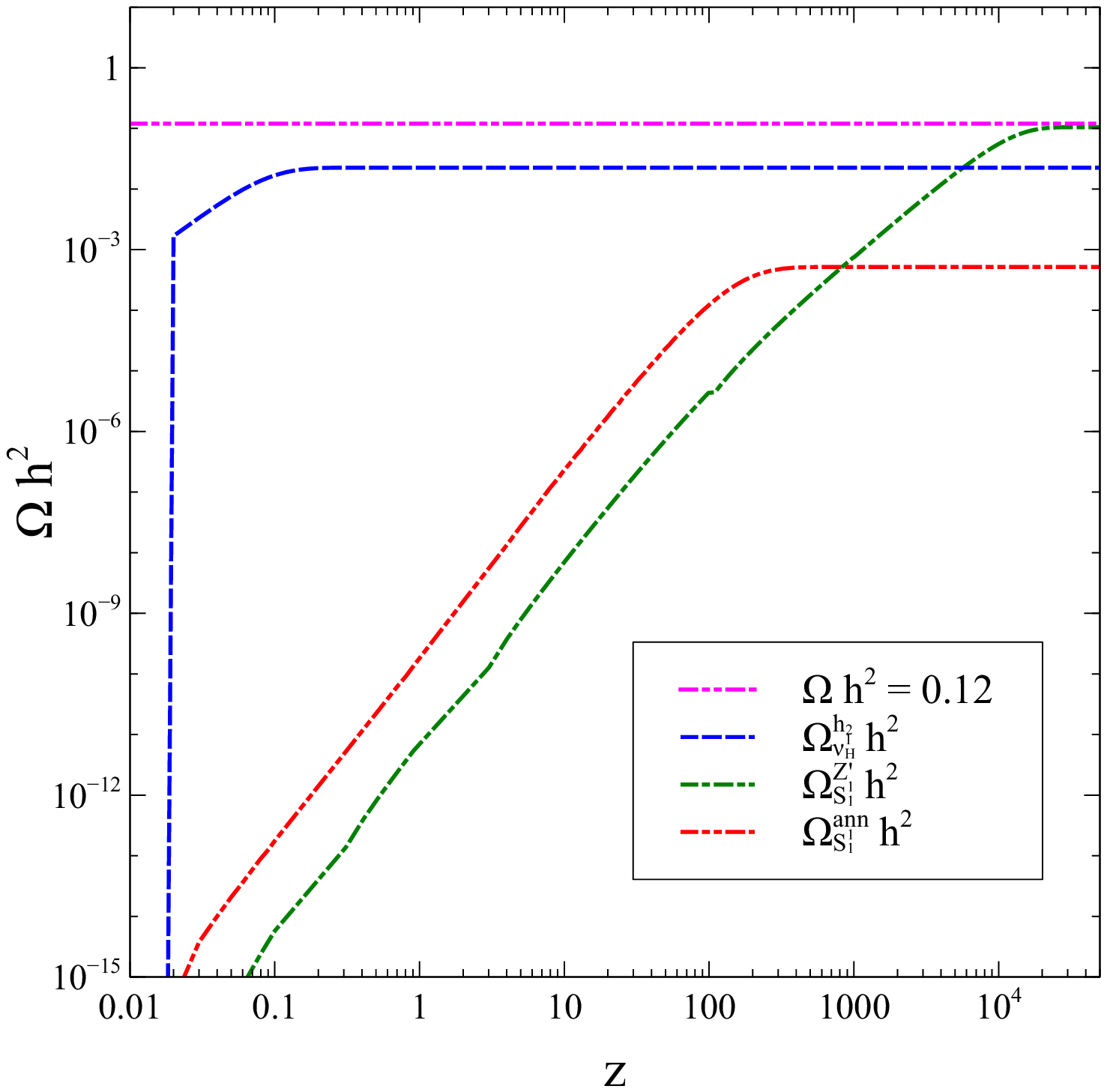}
\caption{Variation of relative relic density contributions of $\nu_H^1$ and $S_1^1$ as a function of 
$z$, for two different sets of model parameters as follows. Left (right) panel:
 $M_{Z'} = 10$~GeV ($2.5$~GeV), $M_{\nu_H^1} = 8$~GeV ($2$~GeV), $M_{S_1^1} = 10$~keV ($100$~keV),
$\alpha = 0.01$~rad, $g' \simeq 2.4 \times 10^{-11}$, $M_{h_2} = 5$~TeV, and $z_0=0.01$.}
\label{relic-hi-zp-n1-s1}
\end{figure}
\subsection{\bf Region II: $M_{Z'} < 2 M_{\nu_H^1}$}
In discussed above, for $M_{Z'}>2M_{\nu_H^1}$ (region~I),  the total relic density is dominated by $Z'$ mediated diagrams. Now we turn to the region~II, where $M_{Z'} < 2 M_{\nu_H^1}$ and $Z'$ decays to $\nu_H^1$ pair is kinematically forbidden, and consequently, $\nu_H^1$ production is $h_2$ dominated. Therefore in region~II, a major portion of our two DM candidates, $\nu_H^1$
and $S_1^1$, is produced almost independently from the $h_2$ and $Z'$ mediated processes, respectively. In other words, by fixing $M_{h_2}$, $g'$ and $M_{\nu^1_H}/M_{Z'}$ at certain values to get a significant contribution from $\Omega_{\nu_H^1}h^2$, one can obtain a relevant $\Omega_{S_1^1}h^2$ contribution independently by changing $M_{S^1_1}$ within keV range. This possibility did not exist in region~I because both $\nu_H^1$ and $S_1^1$ are produced dominantly via $Z'$ and therefore have the same number density. The only way to have comparable contribution from both in region~I would be to raise the mass of $S_1^1$ to the GeV range. However, this is untenable since that will spoil the inverse seesaw mechanism scenario for generating light neutrino masses \cite{MohapatraIS1,Mohapatra:1986bd,Khalil:2010iu,GonzalezGarcia:1988rw}.    
In region~II this lacuna is remedied since here 
$\nu_H^1$ and $S_1^1$ are produced independently - while $\nu_H^1$ are dominantly produced from $h_2$, $S_1^1$ are produced from $Z'$. 

In Fig.~\ref{relic-hi-zp-n1-s1}, for two suitably chosen sets of BLSMIS parameters ($M_{Z'} = 10$~GeV ($2.5$~GeV), $M_{\nu_H^1} = 8$~GeV ($2$~GeV), $M_{S_1^1} = 10$~keV ($100$~keV),
$\alpha = 0.01$~rad, $g' \simeq 2.4 \times 10^{-11}$, $M_{h_2} = 5$~TeV, and $z_0=0.01$), 
we show variation of decay and annihilation contributions to the relic density of $S_1^1$ and $\nu_H^1$, as a function of $z$. In the figure, green dashed-dotted and red dashed-double-dotted lines correspond to the $S_1^1$ relic density contributions ($\Omega^{Z'}_{S_1^1}h^2$ and $\Omega^{\rm ann}_{S_1^1}h^2$), respectively, while
blue dashed line corresponds to $\nu_H^1$ relic density contribution from decay of $h_2$ ($\Omega^{h_2}_{\nu_H^1}h^2$). From the figure, it is clearly seen that $S_1^1$ has a relevant relic density contribution, unlike the situation in region~I. Note that for a larger $S_1^1$ mass ($M_{S_1^1}=100$~keV), the $S_1^1$ contribution to the total relic density even starts to be the dominant one, as seen in the right panel of Fig.~\ref{relic-hi-zp-n1-s1}. 

\section{Conclusion}\label{Con}
In this work we studied two beyond SM problems, {\it viz.}, the non-zero neutrino masses
and the existence of the DM. In studying the tiny neutrino masses, we followed the inverse seesaw
mechanism within the $B-L$ extension of the SM (BLSMIS). Six SM singlet fermions were  introduced for inverse seesaw mechanism to work and three more singlet fermions (with mass of order keV) were added to cancel the $U(1)_{B-L}$ gauge anomaly. The lightest of these additional fermionic states, $S_1^1$, can be a WDM, being odd under a $\mathbb{Z}_2$ discrete symmetry. We studied
$S_1^1$ as a FIMP WDM and showed that it could be produced 
via the freeze-in mechanism from the decay of the extra neutral gauge
boson $Z'$ and the on-shell annihilation processes mediated by $Z'$. 
We showed that the relative contributions to the DM relic density from both the decay and the on-shell
annihilation processes are more or less equal. We scanned over the relevant BLSMIS parameters by imposing the Planck constraint of the 
DM relic density and showed that a large portion of the parameter space gives a small contribution to the DM relic density. Therefore, we studied a two component FIMP DM as a possible scenario in the BLSMIS to get an extra contribution to the DM relic density. In this scenario, the lightest heavy RH neutrino, $\nu^1_H$, can contribute to the DM relic density as an independent DM component (with mass of order GeV). For $M_{Z'}>2M_{\nu_H^1}$, we showed that the production of $\nu_H^1$
as a DM candidate through the $Z'$ mediator has the dominant contribution
to the total DM relic density. On the other hand for 
$M_{Z^{\prime}} < 2 M_{\nu^1_H}$, the $h_2$ mediated processes will contribute dominantly to $\nu_H^1$ production while $Z'$ mediated processes will contribute dominantly to $S_1^1$ production. We emphasized that in this region both FIMP candidates ($S_1^1$ and $\nu_H^1$) can contribute to the total DM relic density. 

\vspace{-0.11cm}
\section*{Acknowledgements}
The authors would like to thank the Department of Atomic Energy
Neutrino Project of Harish-Chandra
Research Institute (HRI). We also acknowledge the HRI cluster
computing facility (http://www.hri.res.in/cluster/).
This project has received funding from the European Union's Horizon
2020 research and innovation programme InvisiblesPlus RISE
under the Marie Sklodowska-Curie
grant  agreement  No.~690575 and  No.~674896. The authors would like to thank Abhass Kumar for fruitful discussions.
     
\appendix
\section*{Appendix}
\section{Analytical expression of the collision terms}
\label{App:AppendixA}
For any generic process $A (\tilde{p}) \rightarrow B(\tilde{p_1})\,
C(\tilde{p_2})$ (where $\tilde{p} = (E_{p},\bar{p})$),
the collision term takes the
following form \cite{Kolb:1990vq, Gondolo:1990dk},
\begin{eqnarray}
\mathcal{C}[f_{A}(p)]&=&\dfrac{1}{2\,E_p}\int
\dfrac{g_B\,d^{3} p_1}{(2\pi)^{3} \,2E_{p_1}}
\dfrac{g_C\,d^{3} p_2}{(2\pi)^{3} \,2E_{p_2}}
(2\pi)^4\,\delta^4(\tilde{p}-\tilde{p_1}-\tilde{p_2})
\times\overline{\,\lvert \mathcal{M}\rvert^2}\nonumber \\
&&~~~~~~~~~~~~~~\times\,[f_B\,f_C\,\left(1 \pm f_A\right)
-f_A\left(1 \pm f_B\right)\left(1 \pm f_C\right)]\,.
\label{colision1}
\end{eqnarray}
From Eq.~(\ref{Z-dis-collision}), we can see that the Boltzmann equation
which determine the distribution function of the extra gauge boson
$Z'$ contains two collision terms one is for its production
and the another one is for its decay. The expression for the
two collision terms are described below.

\begin{itemize}
\item $\mathcal{C}^{Z' \rightarrow {\rm all}}$:
\,\,  
Collision term for the $B-L$ gauge boson decay takes the following form
after using the Eq.~(\ref{colision1}),  
\begin{eqnarray}
\mathcal{C}^{\zmt \rightarrow {\rm all}}&=&-f_{\zmt}(\xi_p,z)\times
\Gamma_{\zmt \rightarrow {\rm all}}\times
\dfrac{r_{\zmt}}{\sqrt{\xi_p^2\,{\mathcal{B}(z)}^2+r_{\zmt}^2}}\,,
\end{eqnarray}
where $r_{Z'} = M_{Z'}/T$ and
$\Gamma_{\zmt \rightarrow {\rm all}} = \Gamma_{\zmt \rightarrow f \bar{f}}$
+ $\Gamma_{\zmt \rightarrow \chi \chi}$. Expression for the each decay terms
are as follows,
\begin{eqnarray}\label{zpdecay}
\Gamma_{\zmt \rightarrow f \bar{f}} &=&
\frac{M_{\zmt} \,n_c (q_f\,g')^{2}}{12\,\pi}
\left( 1 + \frac{2 M_{f}^{2}}{M_{\zmt}^{2}} \right)
\sqrt{1 - \frac{4 M_{f}^{2}}{M_{\zmt}^{2}}}\,, \nn \\ 
\Gamma_{\zmt \rightarrow \chi \chi} &=& \frac{M_{\zmt}
g_{\zmt \chi \chi}^{2}}{24\,\pi}
\left(1 - \frac{4M_{\chi}^{2}}{M_{\zmt}^{2}} \right)^{3/2}\,,
\label{dkz}
\end{eqnarray} 
where $f$ refers to the SM fermions and $\chi=S_1^1,\nu_H^1$.
$n_c$ and $q_f$ are the corresponding color and electric charges, respectively. $ g_{Z' \chi \chi} = g'$
for $S_1^1$ and $g'/2$ for $\nu_H^1$. 

\item $\mathcal{C}^{h_{i} \rightarrow \zmt \zmt}$:\,\,
The expression for this collision term where $h_i$ ($i=1,2$) decays
to $Z'$ pair takes the following form,
\begin{eqnarray}
\mathcal{C}^{h_i \rightarrow \zmt\zmt} &=&
\dfrac{z}{48\pi M_{\rm sc}}\dfrac{g_{h_i\zmt\zmt}^2 \ \mathcal{B}^{-1}(z)}
{\xi_p \sqrt{\xi_p^2{\mathcal{B}(z)}^2+
\left(\dfrac{M_{\zmt}\,z}{M_{\rm sc}}\right)^2}}
\left(2+\dfrac{(M_{h_i}^2-2M_{\zmt}^2)^2}{4M_{\zmt}^4}\right)\nonumber \\ 
&&\times \left(e^{-\sqrt{\left(\xi_{i}^{\rm min}\right)^2
{\mathcal{B}(z)}^2+\left(\frac{M_{h_i}\,z}{M_{\rm sc}}\right)^2}}
\,-\,e^{-\sqrt{\left(\xi_{i}^{\rm max}\right)^2
{\mathcal{B}(z)}^2+\left(\frac{M_{h_i}\,z}{M_{\rm sc}}\right)^2}}
\right) \,,
\label{ch2zblzbl-final}
\end{eqnarray}
where 
\begin{eqnarray}
g_{h_{1}{\zmt} \zmt} &=&
\frac{2 M_{\zmt}^{2} \sin \alpha}{v'},~~~~g_{h_{2}{\zmt} \zmt}=\frac{2 M_{\zmt}^{2} \cos \alpha}{v'}\,, \nn \\[0.1cm]
\xi_i^{\rm min} (\xi_p,z)&=&\dfrac{M_{\rm sc}}{2\,\mathcal{B}(z)\,z\,M_{\zmt}}
\left| \,\eta_i (\xi_p,z)-\dfrac{
M_{h_i}^2 \ \mathcal{B}(z)}{M_{\zmt} \ M_{\rm sc}}\,\xi_p\,z
\right| \,,\nn \\[0.1cm]
\xi_i^{\rm max} (\xi_p,z)&=&\dfrac{M_{\rm sc}}{2\,\mathcal{B}(z)\,z\,M_{\zmt}}
\bigg(\eta_i (\xi_p,z)+\dfrac{M_{h_i}^2 \ \mathcal{B}(z)
}{M_{\zmt} \  M_{\rm sc}}
\,\xi_p\,z \,\bigg)\,,\nn \\[0.1cm]
\eta_i(\xi_p,z)&=& \frac{M_{h_i}\,z}{M_{\rm sc}}
\,\sqrt{\dfrac{M_{h_i}^2}{M_{\zmt}^2}-4}\,\,
\sqrt{\xi_p^2\,{\mathcal{B}(z)}^2+
\left(\frac{M_{\zmt}\,z}{M_{\rm sc}}\right)^2}\,.
\end{eqnarray}
\item Relevant partial decay widths of the scalars $h_i$ ($i=1,2$): 
\begin{eqnarray}
\Gamma_{h_{i} \rightarrow Z' Z'} &=&
\frac{M^3_{h_{i}} g'^2 O^2_i}{32 \pi M^2_{Z'}}\,
\sqrt{1 - \frac{4M^2_{Z'}}{M^2_{h_{i}}}}
\,\left( 1 - \frac{4M^2_{Z'}}{M^2_{h_{i}}}
+ \frac{12 M^4_{Z'}}{M^4_{h_{i}}}\right)\,,
\label{dec-expression-h_i-zz}\\[0.3cm]
\Gamma_{h_{i} \rightarrow \nu_H^1 \nu_H^1} &=&
\frac{ M_{h_i} \ \lambda^2_{\nu^1_H \nu^1_H h_{i}}}{16 \pi}
\left(1 - \frac{4 M^2_{\nu_H^1}}{M^2_{h_i}} \right)^{3/2}\,,
\end{eqnarray}
where $O_1=\sin\alpha$, $O_2=\cos\alpha$, and the coupling $\lambda_{\nu^1_H \nu^1_H h_{i}}$ is defined in Eq.~(\ref{HRHRH-coup}).
\end{itemize}


\end{document}